\documentclass[sigplan,10pt,nonacm]{acmart}
\setcopyright{none}
\acmDOI{}
\acmISBN{}
\hypersetup{keeppdfinfo,pdfpublisher={}}
\makeatletter
\let\hyxmp@parse@acmart\relax
\makeatother

\usepackage{algorithm}
\usepackage{algpseudocode}
\usepackage{subcaption}
\usepackage{placeins}

\newcommand{\sysname}{SlideDP}
\newcommand{\sys}{\textsc{\sysname}}
\AfterEndEnvironment{algorithm}{%
  \ifdefined\slidedpEvaluationFiguresPlaced\else
    \begin{figure*}[t]
    \centering
    \edef\batchscalingplotheight{\the\dimexpr 0.241\textwidth\relax}
    \begin{minipage}[t]{0.36\textwidth}
        \vspace{0pt}
        \centering
        \includegraphics[height=\batchscalingplotheight]{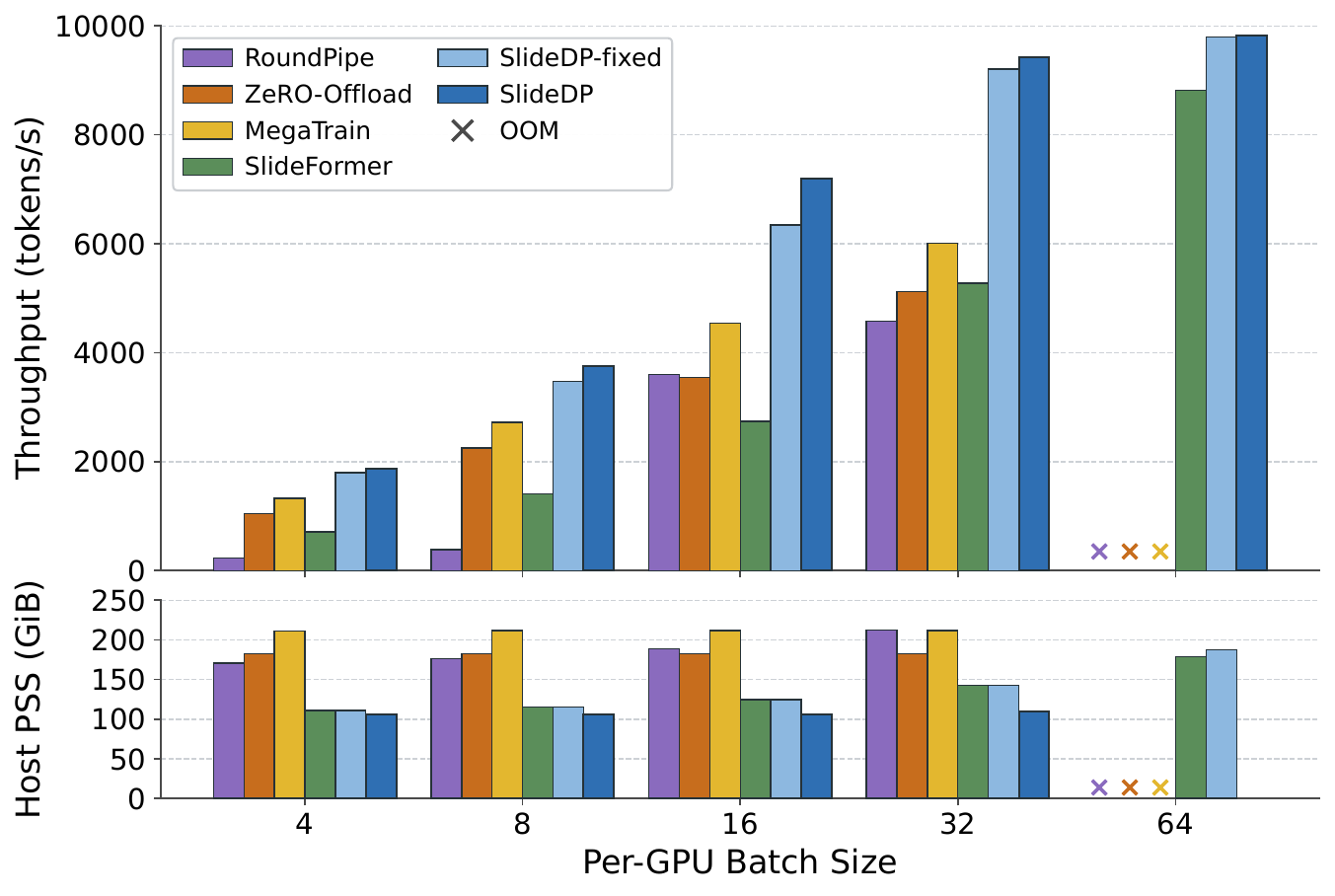}
        \caption{Batch-size scaling and CPU memory for Qwen3-8B on 4$\times$RTX~4090.}
        \label{fig:batch-4090-qwen3-8b}
    \end{minipage}
    \hfill
    \begin{minipage}[t]{0.39\textwidth}
        \vspace{0pt}
        \centering
        \includegraphics[height=\batchscalingplotheight]{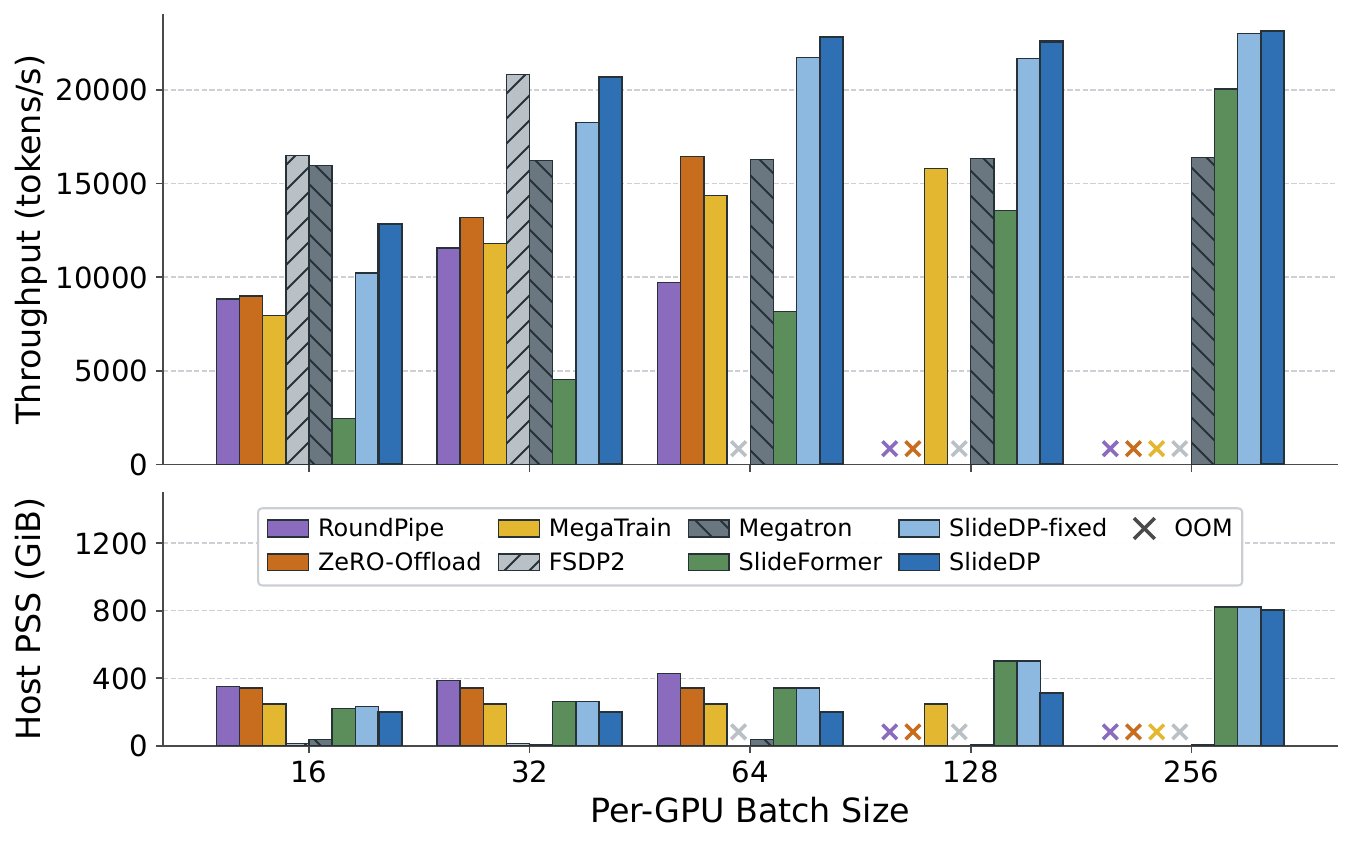}
        \caption{Batch-size scaling and CPU memory for Qwen3-14B on 4$\times$H100.}
        \label{fig:batch-h100-qwen3-14b}
    \end{minipage}
    \hfill
    \begin{minipage}[t]{0.232\textwidth}
        \vspace{0pt}
        \centering
        \includegraphics[height=\batchscalingplotheight]{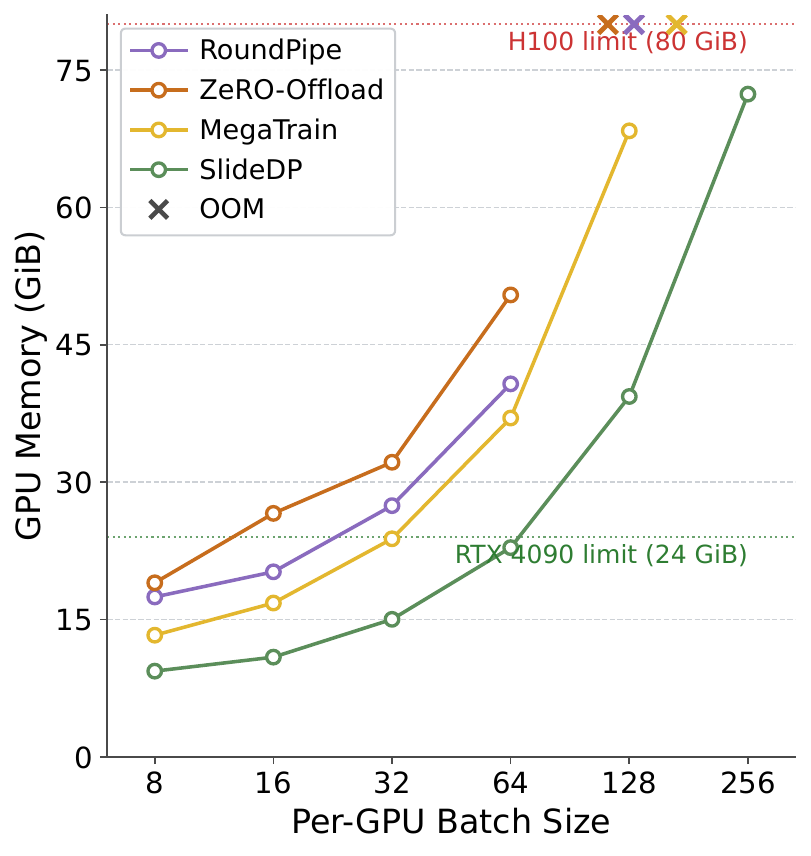}
        \caption{GPU memory vs. batch size for Qwen3-14B.}
        \label{fig:gpu-mem-batch-qwen3-14b}
    \end{minipage}
\end{figure*}
    \global\let\slidedpEvaluationFiguresPlaced\relax
  \fi}

\begin{document}
\title{\sysname: Scaling Host-Resident LLM Fine-Tuning Across Multiple GPUs}
\newcommand{\slidedpHKUSTAffiliation}{%
  \affiliation{%
    \institution{The Hong Kong University of Science and Technology (Guangzhou)}
    \city{Guangzhou}
    \country{China}}}
\newcommand{\slidedpBAAIAffiliation}{%
  \affiliation{%
    \institution{Beijing Academy of Artificial Intelligence}
    \city{Beijing}
    \country{China}}}

\author{Ruijia Yang}
\slidedpHKUSTAffiliation
\email{ryang379@connect.hkust-gz.edu.cn}

\author{Shiyuan Lin}
\slidedpHKUSTAffiliation
\email{slin954@connect.hkust-gz.edu.cn}

\author{Yulong Ao}
\slidedpBAAIAffiliation
\email{aoyulong@outlook.com}

\author{Zhiyu Li}
\slidedpBAAIAffiliation
\email{zyli@baai.ac.cn}

\author{Yingli Zhao}
\slidedpBAAIAffiliation
\email{ylzhao@baai.ac.cn}

\author{Xianduo Li}
\slidedpBAAIAffiliation
\email{xdli@baai.ac.cn}

\author{Yonghua Lin}
\slidedpBAAIAffiliation
\email{ylin@baai.ac.cn}

\author{Zeyi Wen}
\authornote{Corresponding author: \href{mailto:wenzeyi@hkust-gz.edu.cn}{wenzeyi@hkust-gz.edu.cn}.}
\slidedpHKUSTAffiliation
\email{wenzeyi@hkust-gz.edu.cn}
\renewcommand{\shortauthors}{Yang et al.}

\makeatletter
\def\@mkauthors@iii{%
  \gdef\@currentauthors{}%
  \gdef\@currentaffiliation{}%
  \global\setbox\mktitle@bx=\vbox{%
    \unvbox\mktitle@bx
    \centering\normalfont
    {\large
      Ruijia Yang\textsuperscript{1}\quad
      Shiyuan Lin\textsuperscript{1}\quad
      Yulong Ao\textsuperscript{2}\quad
      Zhiyu Li\textsuperscript{2}\par
      Yingli Zhao\textsuperscript{2}\quad
      Xianduo Li\textsuperscript{2}\quad
      Yonghua Lin\textsuperscript{2}\quad
      Zeyi Wen\textsuperscript{1,*}\par}
    \smallskip
    \textsuperscript{1}The Hong Kong University of Science and Technology (Guangzhou)\par
    \textsuperscript{2}Beijing Academy of Artificial Intelligence\par
    \smallskip
    {\small\ttfamily
      \{ryang379,slin954\}@connect.hkust-gz.edu.cn\quad wenzeyi@hkust-gz.edu.cn\par
      aoyulong@outlook.com\quad\{zyli,ylzhao,xdli,ylin\}@baai.ac.cn\par}
    \bigskip}}
\makeatother

\date{}

\begin{abstract}

Host-resident layer streaming enables full-parameter LLM fine-tuning beyond GPU memory, but data-parallel ranks compete for shared host resources. Replicated transfers amplify traffic, while strong scaling can expose host work as computation windows shrink. We present \sys{}, a synchronous data-parallel runtime for shared-host multi-GPU systems. It maintains one authoritative host state, decouples communication routes from state layout, and pipelines parameter delivery, gradient aggregation, and CPU updates across ranks and chunks. An analytical step-time model characterizes resource bottlenecks and pipeline exposure; runtime measurements guide communication, chunking, and activation policies under a GPU memory budget.
In matched-batch sweeps, \sys{} achieves geometric-mean throughput ratios of 1.46--2.64$\times$ over SlideFormer, MegaTrain, and ZeRO-Offload.
On four H100s, \sys{} approaches GPU-resident FSDP2 throughput for Qwen3-14B at a smaller batch size. With a larger batch, it processes over 1M tokens per step and exceeds FSDP2's measured peak throughput by 11.2\%. Separately, it supports 256K-token sequences for the same model and fine-tunes Qwen2.5-72B on four RTX~4090 GPUs.


Project page: \url{https://github.com/RegiaYoung/SlideDP}.
\end{abstract}

\pagestyle{plain}
\maketitle
\hypersetup{pdfauthor={Ruijia Yang, Shiyuan Lin, Yulong Ao, Zhiyu Li, Yingli Zhao, Xianduo Li, Yonghua Lin, Zeyi Wen}}

\section{Introduction}
\label{sec:intro}

\begin{figure}[t]
  \centering
  \includegraphics[width=\columnwidth]{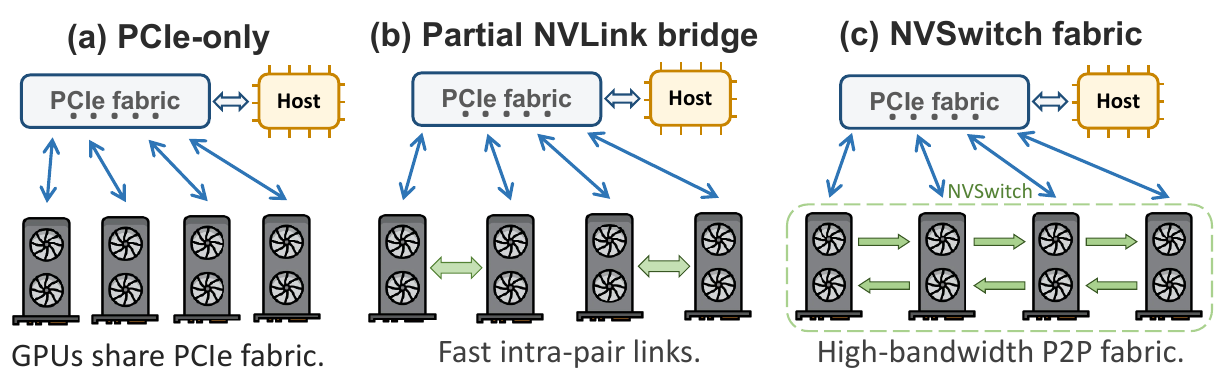}\\
  \includegraphics[width=\columnwidth]{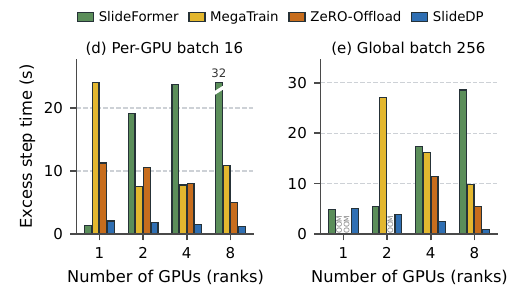}
  \caption{(a--c) GPU interconnects differ, but ranks share host resources. (d,\,e) Excess step time for Qwen3-14B on an eight-A800 node under weak scaling and strong scaling.}
  \label{fig:intro}
\end{figure}

Full-parameter fine-tuning updates every weight in a pretrained LLM, but
mixed-precision Adam maintains substantial model states, including BF16
tensors and FP32 optimizer states. For a model with $N$ parameters, these
states require approximately $16N$ bytes~\cite{rajbhandari2020zero,micikevicius2017mixed},
which can exceed GPU memory capacity or leave insufficient space for
activations even when distributed across GPUs within one node.
Host-resident layer streaming addresses this constraint by keeping
persistent update states in CPU memory and streaming layers through a
small reusable GPU
window~\cite{liao2024lohanlowcosthighperformanceframework,yang2026slideformer,yuan2026megatrain}.
This shifts pressure from GPU memory to host memory and the
CPU--GPU path, making efficiency depend on a \emph{hiding condition}:
host-side parameter delivery, gradient return, and CPU optimizer
updates must overlap with independent GPU computation and complete
before their results are needed.
\sys{} targets multi-GPU training with shared host resources,
coordinating parameter delivery, gradient aggregation, and layer-wise updates across
GPU ranks.

Data parallelism (DP) preserves complete layer computation within each
rank, matching the execution model of layer streaming. Tensor parallelism
partitions operators within layers and introduces inter-GPU
communication~\cite{shoeybi2019megatron}, which can complicate host--device
overlap (\S\ref{sec:background}). A direct DP extension therefore appears
straightforward, but GPUs in one node share a fixed host resource pool.
Increasing active GPUs adds consumers of CPU compute, DRAM bandwidth, and
host--device I/O without proportionally increasing host service capacity.
Shared-PCIe contention has also been studied in offloaded
training~\cite{feng2023mobius}.

Two challenges follow. The first is \emph{avoidable amplification}:
a direct DP extension can replicate each layer transfer across $R$ ranks
and return $R$ gradients for host-side reduction, multiplying demand on
shared host resources. SlideFormer's public multi-GPU implementation
and MegaTrain's multi-GPU mode follow this data-movement
pattern~\cite{yang2026slideformer,yuan2026megatrain}.
The second is \emph{residual pipeline exposure}: even after this
duplication is removed, CPU updates, necessary transfers, and cross-rank
and buffer dependencies remain; under strong scaling, the computation
window available to hide this remaining work can shrink.
Figure~\ref{fig:intro}(d,e) illustrates the resulting execution overhead
under weak and strong scaling. We report \emph{excess step time}
over a per-GPU-batch-matched compute reference, providing an aggregate indicator of exposed communication, host-side work, and runtime coordination overhead.
The design problem is therefore to reduce redundant traffic and schedule
the remaining shared-host work so that it overlaps with GPU computation.

Remaining exposed work depends on both hardware topology
(Figure~\ref{fig:intro}(a--c)) and workload, so there is no single best
communication path. Different routes, such as sharded parameter delivery
or GPU-side gradient reduction, introduce different trade-offs between
communication cost and overlap opportunity. On A800, removing NVLink bridges reverses the preferred delivery route in host-bound workloads, while route differences remain small in GPU-bound workloads (\S\ref{sec:evaluation}). This sensitivity
motivates communication-policy selection adapted to both hardware and
workload.

Partitioned offload already reduces redundant state movement:
ZeRO-Offload and ZeRO-Infinity partition training states across ranks and
reduce unnecessary host--device
transfers~\cite{ren2021zerooffload,rajbhandari2021zeroinfinity}.
Our focus is different: coordinating parameter delivery, gradient return,
and synchronous layer-wise CPU updates within a bounded GPU window when
all ranks share a fixed host resource budget. Reducing logical payload
alone is therefore insufficient; communication routes and execution
schedules must jointly satisfy shared-resource constraints and
parameter-version dependencies.

Two design principles follow. First, persistent host-state layout should be decoupled from parameter delivery and gradient reduction, allowing one authoritative host state while selecting communication routes according to topology and workload. Second, available GPU memory should be allocated according to its effect on exposed pipeline work, since fixed memory policies can be suboptimal across execution regimes. Retaining activations can reduce host--device traffic, while preserving intermediate tensors can reduce recomputation;
their benefits depend on the active bottleneck and pipeline interaction. Communication and activation policies should therefore be guided by shared-resource demands and pipeline dependencies, and validated through measurements.

We present \sys{}, a synchronous DP runtime for full-parameter LLM
fine-tuning on one multi-GPU node. It makes the following three contributions:

\textbf{1. A shared-host runtime for layer-streaming DP.}
We design a runtime that coordinates parameter delivery, gradient reduction and return, and layer-wise CPU updates within a bounded GPU working window. It maintains one authoritative host copy of master weights and optimizer states while supporting replicated or sharded parameter delivery over the same persistent state layout. GPU-side reduction returns one aggregated gradient per layer, while chunking overlaps parameter conversion with transfers and gradient return with CPU updates. The runtime coordinates these operations under per-layer parameter-version and update dependencies to preserve synchronous DP semantics.

\textbf{2. An analytical step-time model of shared-host scaling.}
We model a training step using shared-resource service demands and
cross-layer, cross-rank execution dependencies, including readiness,
buffer reuse, and parameter versions. The model characterizes
host-bound and GPU-bound execution and explains when reducing traffic
shortens a step, when shrinking computation windows expose host work,
and why route choices depend on both topology and workload. We examine
these effects through comparisons of communication routes, GPU counts,
and chunk sizes.

\textbf{3. Pipeline-aware policy selection with measured cost.}
AutoPolicy configures communication routes, chunk sizes, and activation
layouts for the workload and hardware. Elastic Checkpointing supplies
per-layer choices for retention, offloading, and recomputation.
Staged profiling selects routes and chunks, screens activation candidates,
and calibrates their costs in a mini-pipeline before assigning policies
to layers and validating the configuration. We report policy gains
across execution regimes together with selection overhead and the
steps needed to amortize it.

We implement \sys{} in PyTorch Distributed and evaluate it on RTX~4090, A800, and H100 platforms.
Matched-batch sweeps show geometric-mean throughput ratios of
1.46--2.64$\times$ over SlideFormer~\cite{yang2026slideformer},
MegaTrain~\cite{yuan2026megatrain}, and
ZeRO-Offload~\cite{ren2021zerooffload}.
Across 4B--72B models on four H100s, it reaches approximately
85--92\% of compute-only reference throughput projected from
single-GPU measurements.
Qwen3-14B throughput approaches GPU-resident
FSDP2~\cite{zhao2023pytorchfsdp,pytorch2026fsdp2docs} at a smaller batch
and exceeds its measured peak at a larger one.
For Qwen3-32B at 64 sequences per GPU, it maintains 95--98\%
weak-scaling efficiency across two to eight A800s.
On four H100s, \sys{} fine-tunes Qwen3-14B with over 1M tokens
per step without gradient accumulation and, separately, with
256K-token sequences. It also fine-tunes Qwen2.5-72B on a
PCIe-only workstation with four RTX~4090 GPUs.

\section{Background and Motivation}
\label{sec:background}

\subsection{Training States in Full-Parameter Fine-Tuning}
\label{sec:state-anatomy}

Full-parameter fine-tuning stresses the memory hierarchy because each
training step maintains multiple categories of states with different
lifetimes and access patterns. Table~\ref{tab:training-states}
summarizes the major states in mixed-precision Adam
training~\cite{kingma2014adam,micikevicius2017mixed}. For a model with
$N$ parameters, BF16/FP16 compute weights and gradients each require
$2N$ bytes. FP32 master weights and the two FP32 Adam moment buffers
contribute another $4N+8N$ bytes. With memory-efficient
attention~\cite{dao2022flashattention}, activation memory scales as $O(nhsb)$ for fixed architectural ratios,
where $n$, $h$, $s$, and $b$ denote layer count, hidden dimension,
sequence length, and batch size.
Thus, the overall training footprint grows along two major scaling dimensions:
model size increases persistent parameter and optimizer states, while
longer sequences and larger batches increase activation memory.

\begin{table}[h]
\centering
\small
\caption{Training-state memory footprint in mixed-precision full-parameter fine-tuning.}
\label{tab:training-states}
\setlength{\tabcolsep}{4pt}
\begin{tabular}{lccc}
\toprule
State & Precision & Size & Main use \\
\midrule
Compute weights & BF16/FP16 & $2N$ & FWD/BWD \\
Gradients & BF16/FP16 & $2N$ & Sync/update \\
Master weights & FP32 & $4N$ & Update \\
Adam moments & FP32 & $8N$ & Update \\
Activations & BF16/FP16 & $O(nhsb)$ & BWD \\
\midrule
Total estimate & -- & $16N + O(nhsb)$ & Training step \\
\bottomrule
\end{tabular}
\end{table}

Existing techniques reduce individual components of the training
footprint. Activation checkpointing trades recomputation for lower
activation memory~\cite{chen2016training}, while kernel optimizations
reduce temporary memory overhead~\cite{dao2022flashattention,triton,hsu2024ligerkernelefficienttriton}.
Parameter-efficient and optimizer-state methods further reduce
trainable or optimizer memory~\cite{hu2022lora,dettmers2023qlora,peft,NEURIPS2024_2c570b0f,zhao2024galore}. However, these techniques do not determine
how persistent states and transient execution data should be placed and
coordinated across devices, leaving state movement and runtime scheduling
as separate challenges.

Parallelism distributes training computation and states across GPUs,
with different paradigms targeting different execution regimes.
Tensor parallelism partitions operators across GPUs but introduces
communication on computation critical paths~\cite{shoeybi2019megatron}.
Pipeline parallelism partitions layers into stages but requires additional
scheduling and may suffer from pipeline bubbles~\cite{huang2019gpipe,narayanan2019pipedream}.
Data parallelism preserves identical layer computation across ranks while
synchronizing training states. Because DP does not repartition operators
within a layer, it preserves the layer boundaries and local computation
schedule on which host-memory-centric streaming relies.
Therefore, the central question for an
efficient DP runtime is how to own, move, and update the states in
Table~\ref{tab:training-states} under limited GPU memory.

\subsection{GPU-Centric Data Parallelism and Offloading}
\label{sec:gpu-owned-dp}

We use \textit{state ownership} to describe the residency and management
of authoritative training states. Existing DP abstractions remain
rank-centric: states are replicated or partitioned across data-parallel
ranks, with CPU and NVMe serving as additional residency tiers.

\textbf{Replicated and sharded state ownership.}
Distributed Data Parallel (DDP) replicates parameters, gradients, and
optimizer on each GPU rank and synchronizes gradients, without
reducing per-GPU memory~\cite{li2020pytorch}.
ZeRO~\cite{rajbhandari2020zero} and FSDP~\cite{zhao2023pytorchfsdp}
instead partition training states across ranks. Depending on the sharding
level, this reduces optimizer, gradient, and parameter memory but adds
reconstruction and synchronization: parameters are typically materialized
through all-gather before computation, while gradients are synchronized
and resharded through reduce-scatter after backward.

\textbf{Sharded ownership with CPU/NVMe residency.}
Offloading extends this sharded design by moving selected states out of
GPU memory.
ZeRO-Offload moves optimizer states and computation to the
CPU~\cite{ren2021zerooffload}, while ZeRO-Infinity extends the residency
hierarchy to CPU and NVMe~\cite{rajbhandari2021zeroinfinity}.
FSDP also supports CPU offload for sharded states. Other systems further
explore profiling-based state placement and execution policy
selection~\cite{fang2023patrickstar,yang2026protrain}.

When sharding and offloading are combined, relieving the capacity problem
introduces an execution-coordination problem. Sharding introduces GPU-side
collectives for parameter materialization and gradient synchronization
or redistribution, while offloading adds CPU--GPU transfers and
host-side optimizer work. On a multi-GPU node, these operations can
contend for shared PCIe, host-memory bandwidth, and CPU resources while
overlapping with GPU computation. Reducing state redundancy or moving
state to a larger memory tier therefore does not by itself determine
step time; execution also depends on how this work maps onto shared
resources and how much of it remains exposed in the pipeline.

This execution question becomes central once persistent states are
anchored in host memory and GPUs materialize only the states needed for
the current computation, which is the setting we introduce next.

\subsection{Host-Memory-Centric Layer Streaming}
\label{sec:host-memory-centric}

Host-memory-centric systems shift the runtime problem from memory
placement to heterogeneous scheduling. Once GPUs are used as transient
BF16 compute workers over bounded active layers, efficiency depends on
whether host-side optimizer updates, host--device transfers, gradient
movement, and GPU forward/backward computation can be coordinated at a
granularity that hides CPU and PCIe latency.

\textbf{Layer-wise update overlap.}
StrongHold~\cite{sun2022stronghold} shows that a layer's CPU optimizer
update can begin once its gradients are ready and overlap with backward
computation of earlier layers. Its Megatron-based model-parallel
offloading stack, however, does not provide a DP-first shared-host runtime.

\textbf{Host-memory-centric streaming.}
LoHan, SlideFormer, and MegaTrain further explore host-memory-centric
fine-tuning with different emphases~\cite{liao2024lohanlowcosthighperformanceframework,yang2026slideformer,yuan2026megatrain}.
They share the idea that persistent parameters and optimizer states can
reside in host memory while GPUs materialize only tensors required by
active layers. LoHan focuses on activation traffic and recomputation
trade-offs, while SlideFormer and MegaTrain extend the paradigm toward
multi-GPU execution. These advances establish important execution
mechanisms; coordinating them under synchronous DP on a shared host
remains challenging.

\textbf{Pipeline-oriented multi-GPU streaming.}
RoundPipe demonstrates multi-GPU streaming through pipeline parallelism
and CPU offloading~\cite{luo2026roundpipe}. Its overlap of CPU updates
with subsequent computation relies on asynchronous optimizer updates and
delayed parameter versions, allowing the next iteration to proceed before
CPU updates finish. Standard synchronous fine-tuning restores the
optimizer-update dependency and can reintroduce stalls.

Extending host-memory-centric layer streaming to multi-GPU DP requires
more than running multiple single-GPU instances. The challenge is
coordinating multiple GPU ranks that share one authoritative host state
under synchronous data-parallel semantics. A DP runtime must jointly
manage parameter delivery, gradient aggregation, CPU updates, and
execution dependencies across ranks. The following observations
characterize how shared resources and overlap opportunities shape their
costs.

\subsection{Data Parallelism Is Not Free on a Shared Host}
\label{sec:shared-host}

Single-GPU host-memory-centric systems make \emph{one} GPU efficient by
hiding host work behind that GPU's computation. Moving from one GPU to $R$
GPUs creates two primary pressures on this hiding condition: avoidable
host demand can grow with $R$, while strong scaling can shrink the
computation window available to hide the remaining work. Neither effect
is captured by logical byte counts alone: traffic maps onto physical paths
with finite service rates, and pipeline dependencies determine which
service becomes exposed in step time.
We make four observations along this chain;
\S\ref{sec:design} turns them into a design and \S\ref{sec:evaluation}
measures them.

\begin{table}[t]
\centering
\small
\caption{Host traffic per iteration for a streamed layer under four
datapaths over one shared host state. $P_\ell$/$G_\ell$: parameter and
gradient bytes; $q_\ell$: parameter loads per iteration; $R$: rank count.
Sharded delivery adds an all-gather per load. Padding and replicated
tied-weight tails are excluded.}
\label{tab:host-traffic}
\setlength{\tabcolsep}{4pt}
\begin{tabular}{llccc}
\toprule
Delivery & Reduction & H2D & D2H & Host work \\
\midrule
replicated & CPU & $R q_\ell P_\ell$ & $R G_\ell$ & reduce + 1 update \\
replicated & GPU & $R q_\ell P_\ell$ & $G_\ell$ & 1 update \\
sharded & CPU & $q_\ell P_\ell$ & $R G_\ell$ & reduce + 1 update \\
sharded & GPU & $q_\ell P_\ell$ & $G_\ell$ & 1 update \\
\bottomrule
\end{tabular}
\end{table}

\begin{figure}[t]
  \centering
  \includegraphics[width=\columnwidth]{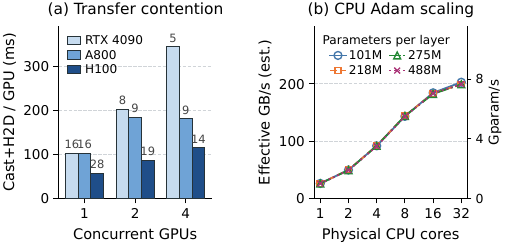}
  \caption{Shared-host microbenchmarks.
  (a) Per-GPU cast+H2D for an 800M-parameter layer under concurrent GPUs.
  (b) CPU Adam core number scaling on Xeon 8468.}
  \label{fig:shared-host-microbench}
\end{figure}

\textbf{Observation 1---Avoidable amplification.}
A direct DP extension causes host demand to grow with GPU rank count,
while host service capacity remains fixed.
Table~\ref{tab:host-traffic} accounts for the per-layer
traffic of the four datapaths a DP runtime can choose from. A direct
extension of a single-GPU engine---replicated delivery and CPU-side
reduction---multiplies both directions by $R$ and adds a host-side
reduction that competes with the optimizer update for DRAM bandwidth.
These resources are shared within the host/NUMA domain rather than
provisioned independently for each rank.
On an A800 node, the cast-and-copy of an 800M-parameter layer
sustains 15.6\,GB/s per GPU when one GPU copies but 8.7\,GB/s per GPU
when two copy concurrently (Figure~\ref{fig:shared-host-microbench}a).
On the H100 host, CPU Adam reaches an estimated 182--184\,GB/s at
16 physical cores across 101--488M-parameter layers. Increasing to
32 cores improves update speed by only 8.5--10.0\%
(Figure~\ref{fig:shared-host-microbench}b).
These diminishing returns complement the transfer results: increasing
concurrent users or CPU workers does not provide proportional growth in
shared-host service capacity, making rank-dependent host amplification a
scaling concern.
Removing avoidable amplification still leaves host work that the
pipeline must hide.

\textbf{Observation 2---Residual pipeline exposure.}
Host-side work becomes exposed when its completion exceeds the available
hiding window before downstream operations, such as buffer reuse or the
next parameter dispatch, require the result.
The window may include remaining backward computation and early
computation in the next iteration.
For example, under fixed-global-batch strong scaling, increasing $R$
reduces each rank's batch and can shrink this window, although compute
time need not scale exactly as $1/R$.
Exposure depends on shared-host service capacity, queueing, and cross-rank
readiness, as captured by the step-time model
(Section~\ref{sec:step-time-model}).
Figure~\ref{fig:intro}(e) illustrates this scaling challenge.
For Qwen3-14B at a fixed global batch of 256, increasing the A800
GPU count from four to eight reduces the matched compute reference
from 27.2 to 14.4~s, while SlideFormer's step time remains near
43~s. \sys{} reduces step time from 29.6 to 15.3~s, demonstrating
effective overlap as the per-rank computation window shrinks.
Under weak scaling, the local computation window is largely preserved,
but more ranks can still increase shared-host contention.
Configurations dominated by exposed host/transfer work are host-bound;
those dominated by accelerator computation are GPU-bound.
Exposed cost therefore depends not only on remaining work, but also on
how execution routes map it onto shared resources.

\textbf{Observation 3: the right delivery route depends on topology
\emph{and} regime.} Sharded delivery cuts each rank's parameter payload
to $P_\ell/R$ but adds an all-gather on the GPU interconnect. On a
4$\times$A800 node with NVLink bridges and Qwen3-14B at 16 sequences
per GPU, sharded delivery is 17\% faster than replicated; after the
bridges are removed, replicated is 26\% faster. At 8
sequences per GPU the gaps are 33\% and 19\%. In GPU-bound eight-GPU workloads
(64--256 sequences per GPU), the two routes differ by under 2.5\%
(\S\ref{sec:evaluation}). A wrong route matters most when transfers are
exposed, which is why the choice cannot be made from topology alone.
GPU memory allocation likewise shapes the remaining pipeline exposure.

\textbf{Observation 4---The value of spare HBM depends on exposed pipeline work.}
Bounded layer streaming can leave substantial HBM headroom. At 32
sequences per GPU, the fixed Qwen3-14B configuration in
Table~\ref{tab:autopolicy-ablation} uses 15.0\,GiB of peak CUDA reserved
memory on an 80-GB H100. This headroom can reduce different sources of
exposed work: retaining saved activations on the GPU removes offload and
reload traffic, whereas retaining additional intermediate results reduces
recomputation. Because these choices reduce different components of the
execution pipeline, their benefit depends on which component is exposed.
Because spare HBM can reduce different exposed costs depending on the
active bottleneck, memory allocation itself becomes a runtime policy
decision rather than a fixed checkpointing choice.
Fixed memory policies can therefore be suboptimal across regimes.

Together, these observations show that shared-host DP requires
coordinated decisions across state movement, execution scheduling, and
memory allocation rather than independent optimization of individual
components.


\section{Design of \sys{}}
\label{sec:design}

\subsection{Overview}
\label{sec:design-overview}
\sys{} is a shared-host runtime for synchronous data-parallel training
on a multi-GPU node. It coordinates $R$ GPU ranks through one
authoritative host state and a cross-rank layer pipeline. Host memory
holds the FP32 master parameters and Adam states; each rank processes
its local data using temporary BF16 layer copies in a bounded, reusable
GPU window. The runtime jointly schedules parameter delivery, gradient
aggregation, buffer reuse, and layer updates to overlap host work with
GPU computation while preserving synchronous DP semantics.

\begin{figure*}[t]
  \centering
  \includegraphics[width=\textwidth]{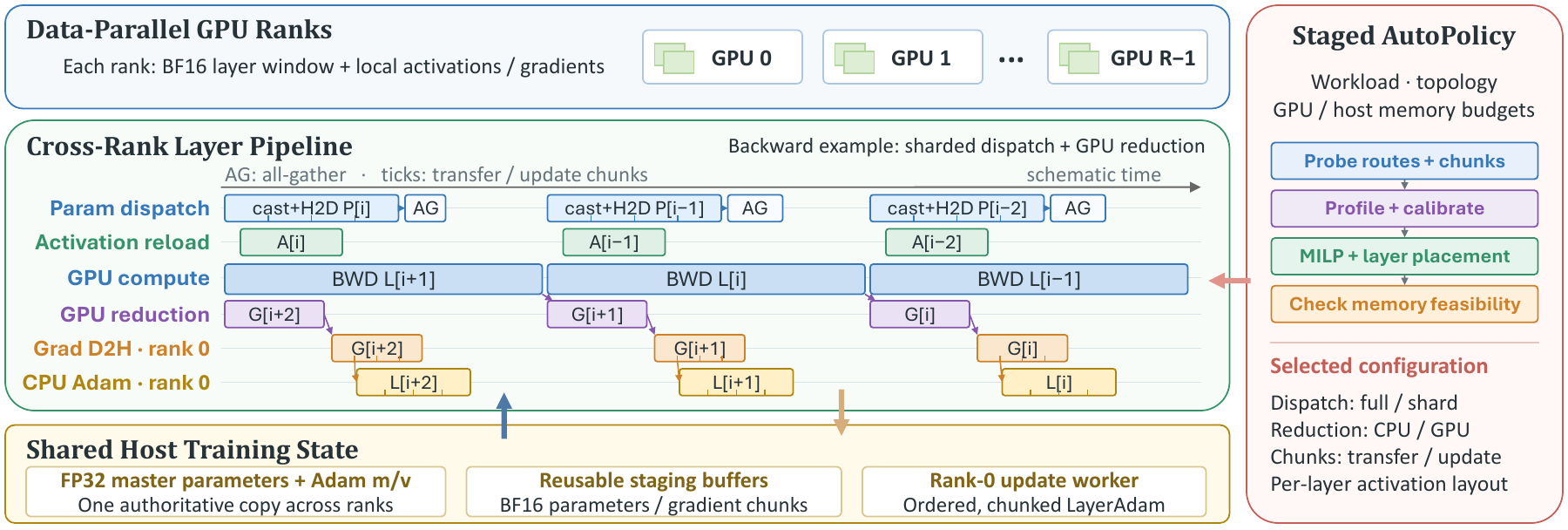}
  \caption{Overview of \sys{}: shared host state, cross-rank pipelining, and AutoPolicy.}
  \label{fig:slidedp-overview}
\end{figure*}

Figure~\ref{fig:slidedp-overview} connects these mechanisms to their
policy choices. Shared state and communication routes define what work
is performed (Section~\ref{sec:state-model}); the cross-rank pipeline
coordinates its execution and overlap (Section~\ref{sec:pipeline}).
The model then explains how resource demands and dependencies shape
scaling (Section~\ref{sec:step-time-model}). Elastic Checkpointing
exposes activation retention, offloading, and recomputation choices
(Section~\ref{sec:elastic-checkpointing}); AutoPolicy measures their
pipeline effects and selects a memory-feasible configuration
(Section~\ref{sec:autopolicy}).

\subsection{Shared Host State and Communication Routes}
\label{sec:state-model}
\label{sec:transfer-strategy}
\sys{} separates persistent state ownership from communication routes.
For each of $L$ layers, the host holds one shared copy of FP32 parameters
$\Theta_l^{32}$ and Adam moments $(m_l,v_l)$. The rank-0 optimizer worker
updates this state once per iteration. Three invariants preserve
synchronous DP: all ranks compute a layer with the same parameter
version; its update consumes the gradient aggregated across all ranks;
and its next-iteration dispatch waits for that update. GPU windows hold
temporary copies, so changing the delivery route does not change state
ownership or update semantics.

\textbf{Parameter dispatch.}
For a streamed layer with $P_l$ BF16 parameter bytes, full dispatch
sends $P_l$ to each rank; sharded dispatch sends $P_l/R$ to each rank
and reconstructs the layer with all-gather. If the layer is loaded
$q_l$ times per iteration, aggregate H2D traffic is $Rq_lP_l$ or
$q_lP_l$, respectively, ignoring padding. Most layers reload parameters for
backward after forward copies leave the window; $q_l$ counts actual loads. Replicated tied-weight tails are accounted for
separately. Both routes leave the host master state unsharded.

\textbf{Gradient return.}
For a $G_l$-byte communicated gradient, GPU reduction sums rank-local
gradients to rank~0, which averages and returns one gradient to the
host. CPU reduction returns all local gradients, using $RG_l$ D2H bytes
before host-side aggregation. Both paths perform one optimizer update.
For sharded layers with fixed $q_l$, sharded dispatch with GPU reduction
therefore keeps aggregate parameter and gradient payload independent of
$R$. Activations, collectives, and replicated exceptions contribute
additional resource demands.

\textbf{Routes and physical resources.}
The best route depends on the resources carrying its traffic. Parameter
shards traverse shared host paths, while reconstruction uses GPU
interconnects. NVLink or NVSwitch can make
all-gather cheaper than repeated H2D; a slower collective path can favor
full dispatch. CPU reduction also competes with Adam for host resources.
The runtime selects delivery and reduction paths for these shared
resources; Sections~\ref{sec:eval-runtime} and~\ref{sec:eval-scaling}
evaluate demand reduction and delivery-route sensitivity. Control threads, the
update worker, and pinned buffers retain the same CPU/NUMA placement
during profiling and execution.

\subsection{Cross-Rank Pipelining and Chunked Overlap}
\label{sec:pipeline}
\sys{} overlaps work at two scales: completed layers release updates
while earlier layers continue backward, and chunks overlap preparation,
movement, and update within a layer. Six logical lanes coordinate
parameter preparation and H2D, activation offload and reload, GPU compute,
gradient reduction, gradient D2H, and CPU LayerAdam. These lanes share
DRAM and host-device paths. The runtime coordinates rank readiness and
buffer completion so that each layer advances through these lanes with
one aggregated gradient and one host update.

\textbf{Layer order and rank readiness.}
During forward, parameter preparation for the next layer overlaps with
current-layer computation. Backward visits layers in reverse order,
reloading parameters and offloaded activation state before use.
Figure~\ref{fig:slidedp-overview} shows the GPU-reduction path for each layer:
\[
 \{\mathrm{BWD}_{l,r}\}_{r=0}^{R-1}
 \longrightarrow\mathrm{Reduce}_l
 \longrightarrow\mathrm{D2H}_{l,k}
 \longrightarrow\mathrm{Adam}_{l,k},
\]
where $k$ indexes gradient chunks. Once reduction combines all ranks'
contributions, rank~0 returns the aggregated gradient and releases its
chunks to the host update worker; the other ranks participate in
completion synchronization. GPU computation continues on earlier layers
while the completed layer is updated. This moves CPU optimizer work
out of the step-end serial tail while preserving one update per layer.

\textbf{Completion and buffer reuse.}
A GPU slot becomes reusable after its outstanding computation,
collective, and copy operations complete. Host gradient staging likewise
waits for its previous update consumer. One update worker consumes
layers in queue order and their chunks as they become ready. Completion
signals release buffers and publish the updated layer version.
At iteration $t+1$, every rank's dispatch of layer $l$ waits for that
layer's update from iteration $t$. This per-layer gate preserves the
same parameter view across ranks while allowing independent operations
to proceed as their inputs and buffers become ready. Together, version
gates and buffer completion bound in-flight storage and preserve
synchronous DP throughout the overlapped execution. When buffers fill,
backpressure exposes the pending transfer or update work.

\textbf{Overlap within a layer.}
On parameter delivery, the CPU casts chunk $k+1$ to BF16 while DMA
transfers chunk $k$. Aligned AVX-512 staging uses non-temporal stores
to reduce cache pollution from buffers consumed by DMA. All-gather
starts after the complete local shard is ready. On gradient return,
whole-layer reduction precedes chunked D2H; each completed copy releases
its Adam chunk while later copies continue. These two overlap paths
shorten the delay before dependent work can start. Smaller chunks also
incur more launches and queue operations, so AutoPolicy measures the
tradeoff under the chosen routes.

\textbf{Exposed waiting.}
Work is hidden when it finishes before its consumer would otherwise be
ready. For a layer update, consumers include buffer reuse and the next
parameter dispatch, which may follow computation in the next iteration.
Thus the available overlap window follows each operation's next
consumer across the layer pipeline. Section~\ref{sec:step-time-model}
models these dependencies together with shared-resource demand.
Section~\ref{sec:eval-runtime} evaluates chunking with communication
routes and activation layout held fixed.

\subsection{Modeling Shared-Host Scaling}
\label{sec:step-time-model}
The same logical payload can incur different delays depending on its
physical route, concurrent work, and readiness constraints. We model
these effects through resource service demands and operation
dependencies. Inputs are the workload, communication routes, chunk sizes,
activation layout, and resource service costs; the constraints follow
the runtime in Sections~\ref{sec:state-model}--\ref{sec:pipeline}.

\textbf{Resource demand.}
Let $W_j$ be the work per iteration assigned to resource $j$ and
$\beta_j^{\max}$ an upper bound on its service rate. Capacity gives the
steady-state lower bound
\[
 T_{\mathrm{step}}\geq\max_j D_j^{\min},\qquad
 D_j^{\min}=W_j/\beta_j^{\max}.
\]
Resources include GPU compute and copy engines, GPU interconnects, the
CPU update worker, host DRAM, and shared PCIe paths. Demands from ranks
using a shared path are summed against that path's capacity; independent
GPU demands remain separate. Measured effective rates give empirical
estimates $\widehat D_j=W_j/\widehat\beta_j$. Concurrent casting, DMA,
and Adam affect these rates. Combined probes such as cast+H2D contribute
one joint service cost.

\textbf{Execution dependencies.}
For a fixed order of compute, collective, and transfer/update operations,
completion obeys
\[
 f(v)=d_v+\max\bigl(\{a_v\}\cup\{f(u):u\in\mathcal P(v)\}\bigr).
\]
Here $a_v$ is the release time, $d_v$ excludes predecessor waits, and
$\mathcal P(v)=\mathcal P_G(v)\cup\mathcal P_R(v)\cup\mathcal P_B(v)$.
The three sets encode data and parameter-version dependencies, ordering
on serialized streams or workers, and buffer-release dependencies,
respectively. Collective readiness includes all participating ranks.
Shared bandwidth changes service times rather than imposing one serial
order on all CPU and DMA operations.

For example, a layer's backward computation waits for its parameters,
required activations, and the preceding GPU computation. Sharded
delivery can advance parameter readiness, but only after all-gather
reconstructs the layer. This advances backward execution only while
parameter readiness is the latest predecessor; activation reload or
GPU computation can instead determine its start. Buffer availability
and the prior update further constrain when delivery can begin.
The model thus identifies which readiness constraint a route change
must advance to accelerate backward execution.

The graph extends across iterations: the next dispatch of a layer waits
for its update. Steady-state step time is the interval between successive
iteration boundaries after warmup. The final update drain is a separate
completion interval, rather than a cost charged in full to every step.

\textbf{Regimes and scaling.}
When compute and host/transfer service dominate, their interaction has
the coarse approximation $T_{\mathrm{step}}\approx\max\{C,H\}+E$.
$C$ includes forward, backward, and replay computation; $H$ is the
bottleneck in host and transfer service. $E$ accounts for additional
delay caused by execution dependencies and pipeline boundaries.
The dependency model captures buffer waiting and rank readiness.

In the host-bound regime ($H>C$), reducing $H$ by $\delta H\geq0$
saves $\min(\delta H,H-C)$ within this approximation when $C$ and $E$
remain fixed. Further host savings become hidden once compute dominates.
Reducing replay instead lowers $C$ and can help GPU-bound execution.
A route or activation change can affect both demands and readiness;
AutoPolicy therefore measures its effect in the pipeline.

Strong scaling gives
$C(R)=C(B_{\rm global}/R,S,p_{\rm ckpt})$, where $S$ is sequence length
and $p_{\rm ckpt}$ the activation layout.
If compute efficiency is stable, $H$ nearly constant, and boundary costs
small, $C(R)\approx C(1)/R$ gives a crossover
$R_{\rm cross}\approx C(1)/H$. Under weak scaling, sharding can shorten
per-rank H2D service at fixed local batch. Aggregate throughput can then
grow superlinearly while H2D limits the step, until shared host capacity
or all-gather takes over.
Section~\ref{sec:eval-scaling} examines these implications through GPU-count
scaling and route comparisons across computation windows and topologies.

\subsection{Elastic Checkpointing}
\label{sec:elastic-checkpointing}
Elastic Checkpointing defines the activation-side policy space evaluated
by AutoPolicy. It uses memory beyond the bounded GPU parameter window
to reduce activation movement or replay computation, combining
checkpoint granularity with activation placement, as explored in
hybrid-parallel training~\cite{yuan2024efficientactivation}.
A layout $p_{\rm ckpt}=(\pi_1,\ldots,\pi_L)$ specifies each layer's
checkpoint granularity, activation retention, and offload ratio.
Checkpoint families and boundary masks or operator presets determine
which tensors are saved and what is replayed. Retaining previously
offloaded activations on GPU removes their D2H/H2D demand and reload
dependencies; recomputation lowers memory demand but adds computation.
Saving additional intermediates reduces replay. Candidate policies trade
GPU memory usage against transfer demand, recomputation cost, and exposed
execution work.

\begin{figure*}[t]
  \centering
  \begin{minipage}[t]{0.48\textwidth}
    \vspace{0pt}
    \centering
    \includegraphics[width=\linewidth]{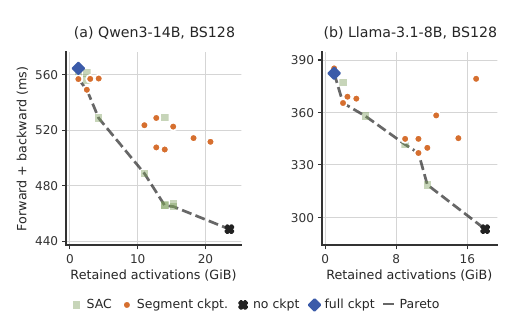}
    \caption{Single-layer checkpointing Pareto frontiers for both models
    (S1024; H100 in (a)).}
    \label{fig:elastic-ckpt-pareto}
  \end{minipage}\hfill
  \begin{minipage}[t]{0.48\textwidth}
    \vspace{0pt}
    \centering
    \includegraphics[width=\linewidth]{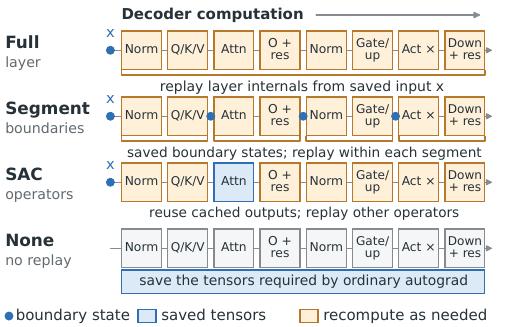}
    \caption{Checkpoint granularities and recomputation
    within one decoder layer.}
    \label{fig:activation-policy-semantics}
  \end{minipage}
\end{figure*}

\textbf{Placement and granularity.}
Figure~\ref{fig:activation-policy-semantics} separates checkpointing
families. Each combines with an offload ratio $r\in[0,1]$ that controls
activation placement, targeting a fraction $r$ of saved bytes moved to
CPU at tensor granularity. The ratio trades GPU residency for CPU--GPU
transfer demand without changing the chosen recomputation policy, and
applies across full-layer, segment-level, operator-level, and
no-checkpoint policies.

Full-layer checkpointing reconstructs intermediates from saved layer
inputs~\cite{chen2016training}. Segment-level checkpointing uses seven-bit
\emph{checkpoint-boundary masks} to define finer replay boundaries,
retaining segment inputs and cross-boundary state needed to reconstruct
intermediates. A compact mask set exposes meaningful boundaries in the
fused decoder. Both use non-reentrant checkpointing, stopping replay once
required tensors are recovered~\cite{pytorch_checkpoint_docs}.
Selective activation checkpointing (SAC) caches chosen operator outputs
and recomputes other operations~\cite{pytorch_checkpoint_docs}.
Segment masks and SAC offer complementary choices over execution
boundaries and operator outputs, varying retained memory, replay scope,
and transient tensor lifetimes. No-checkpoint policies use ordinary
autograd without replay.

\textbf{Measured candidates.}
AutoPolicy instantiates a finite set of masks, operator presets, and
offload ratios, merging ratios that select the same saved tensors within
a checkpoint configuration.
Figure~\ref{fig:elastic-ckpt-pareto} shows measured single-layer
memory--compute tradeoffs for two models, identifying Pareto-efficient
choices at different retained-memory budgets.
AutoPolicy selects candidates based on measured pipeline effects rather
than memory reduction alone. It calibrates frontier and reference
candidates in the runtime, accounting for transient replay peaks before
allocating layers.

\textbf{Layout across layers.}
Static tensor specifications identify saved tensors moved to pinned host
buffers, with cross-layer reload prefetch. The planner groups offloaded
layers contiguously and places GPU-retained policies toward the decoder
tail, ordered from lighter to heavier memory use. Backward reaches this
tail first, releasing large retained states early. Selected layouts must
preserve pipeline feasibility; final placement and memory requirements
are validated in the complete runtime.

\subsection{Measurement-Guided Policy Selection}
\label{sec:autopolicy}
AutoPolicy selects
$\Pi=(p_{\rm param},p_{\rm red},c_{\rm H2D},c_{\rm D2H},p_{\rm ckpt})$
for a workload and topology. The analytical model identifies how these
choices affect demand and readiness; execution measurements supply the
selector's costs. Routes and chunks are selected first, then activation
choices are calibrated under that configuration. This staging controls
selection overhead and captures their interaction with parameter and gradient
transfers. Algorithm~\ref{alg:autopolicy} summarizes the stages.

\textbf{Measure pipeline effects.}
Route probes compare parameter delivery and gradient return; chunk
probes measure their preparation/update overlap. Single-layer profiling
records computation time, retained activations, offload traffic, and
peak-memory metadata. A reduced real training pipeline then replaces
$k_j$ reference layers with each shortlisted candidate $j$.
Let $T_j(k_j)$ and $T_0$ denote the maximum per-rank mean steady-state
step times of the candidate and matching reference pipelines.
With common reference-layer cost $\tau_0$, the calibrated coefficient is
\[
 \tau_j=\tau_0+\frac{T_j(k_j)-T_0}{k_j},
\]
The measured difference captures marginal pipeline cost (overlap and contention). Since total
layer count is fixed, $\tau_0$ does not affect the allocation ranking.
Larger-span probes refine selected candidates before final allocation.
Memory calibration separates per-layer residency, shared prefetch pools,
and transient headroom, with guards across ranks.

\begin{algorithm}[t]
\small
\caption{Measurement-guided AutoPolicy in \sys{}}
\label{alg:autopolicy}
\begin{algorithmic}[1]
\Require Workload $\mathcal W$; topology $\mathcal T$; memory budgets $B$
\Ensure Memory-validated policy $\Pi$
\State $(p_{\rm param},p_{\rm red})\gets\textsc{ProbeRoutes}(\mathcal W,\mathcal T)$
\State $(c_{\rm H2D},c_{\rm D2H})\gets\textsc{TuneChunks}(\mathcal W,p_{\rm param},p_{\rm red})$
\State $\rho\gets(p_{\rm param},p_{\rm red},c_{\rm H2D},c_{\rm D2H})$
\State $\mathcal C_0\gets\textsc{ProfileCkpt}(\mathcal W)$
\State $\mathcal C\gets\textsc{ShortlistAndProbe}(\mathcal C_0,\rho)$
\State $n\gets\textsc{SolveMILP}(\mathcal C,B)$
\State $(\mathcal C,B)\gets\textsc{CalibrateSelectedSpans}(\mathcal C,n,B)$
\State $p_{\rm ckpt}\gets\textsc{PlaceLayers}(\textsc{SolveMILP}(\mathcal C,B))$
\State $\Pi\gets\textsc{ValidateAndRefine}(\rho,p_{\rm ckpt},\mathcal C,B)$
\State \Return $\Pi$ or a feasible conservative plan
\end{algorithmic}
\end{algorithm}

\textbf{Allocate and place.}
The MILP minimizes $\sum_j n_j\tau_j$, where $n_j$ is the integer
layer count of candidate $j$ and $\sum_j n_j=L$.
Relative to the same measured reference, $m_j$, $p_j$, and $h_j$ denote
increments in per-layer residency, shared-pool memory, and transient
headroom. For selected candidates $\mathcal J_+=\{j:n_j>0\}$,
the GPU constraint is
\[
 \sum_j n_jm_j+\max_{j\in\mathcal J_+}p_j
 +\max_{j\in\mathcal J_+}h_j\leq B_{\rm avail}.
\]
$B_{\rm avail}$ subtracts the reference peak reserved memory and a
safety margin from GPU capacity. Shared pools and transient headroom
are thus reserved once at their maximum requirements.
Any configured host-memory and activation-transfer budgets add
constraints; candidate count limits bound extrapolation from measured
spans. The placement rule above converts counts into $p_{\rm ckpt}$.
The objective is a calibrated cost surrogate under the selected routes
and chunks.

\textbf{Validation and selection overhead.}
Short full-model runs validate memory feasibility and compare step
times. Reserved-memory feedback adjusts the budget for a bounded number
of re-solves, retaining the fastest validated feasible configuration.
Formal throughput runs then assess the selected policy's performance.
Transfer and activation measurements can be cached with matching
hardware, workload, and implementation identities. With selection
overhead $W$ and per-step saving $\Delta t>0$, the break-even count is
$\lceil W/\Delta t\rceil$. Section~\ref{sec:eval-policy} reports measured
selection overhead and amortization alongside the gains over the fixed
policy.

\ifdefined\slidedpEvaluationFiguresPlaced
\else
\begin{figure*}[t]
    \centering
    \edef\batchscalingplotheight{\the\dimexpr 0.241\textwidth\relax}
    \begin{minipage}[t]{0.36\textwidth}
        \vspace{0pt}
        \centering
        \includegraphics[height=\batchscalingplotheight]{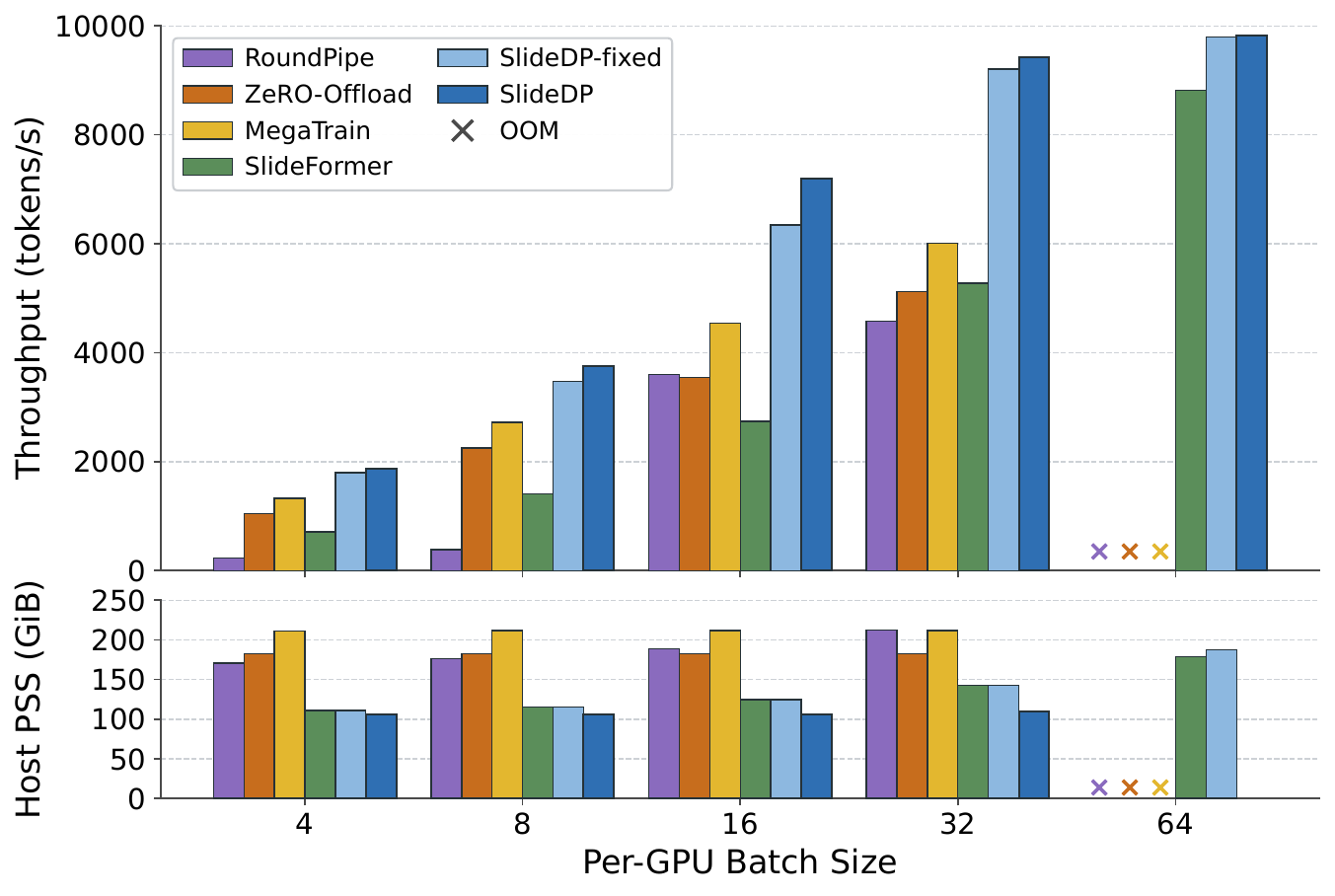}
        \caption{Batch-size scaling and CPU memory for Qwen3-8B on 4$\times$RTX~4090.}
        \label{fig:batch-4090-qwen3-8b}
    \end{minipage}
    \hfill
    \begin{minipage}[t]{0.39\textwidth}
        \vspace{0pt}
        \centering
        \includegraphics[height=\batchscalingplotheight]{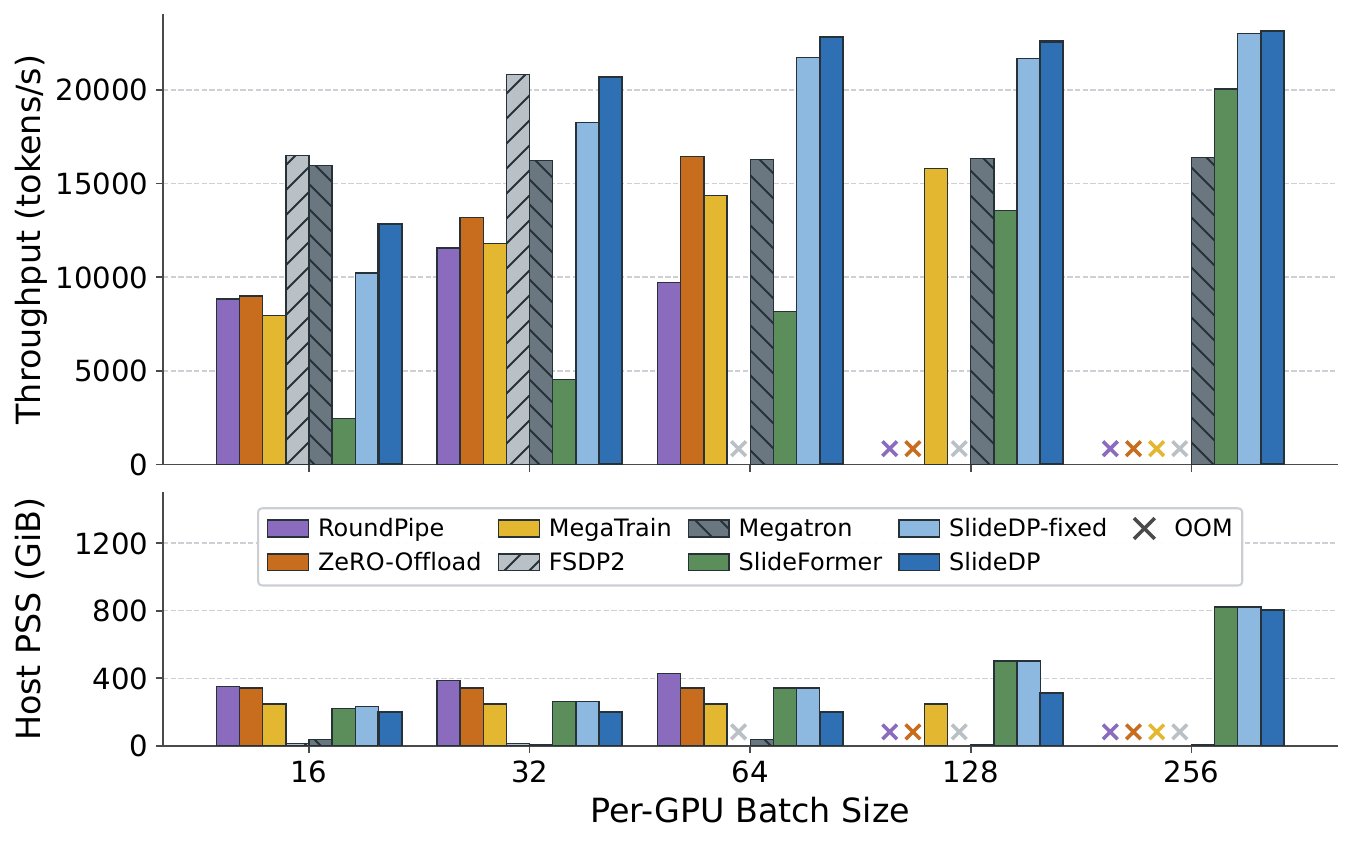}
        \caption{Batch-size scaling and CPU memory for Qwen3-14B on 4$\times$H100.}
        \label{fig:batch-h100-qwen3-14b}
    \end{minipage}
    \hfill
    \begin{minipage}[t]{0.232\textwidth}
        \vspace{0pt}
        \centering
        \includegraphics[height=\batchscalingplotheight]{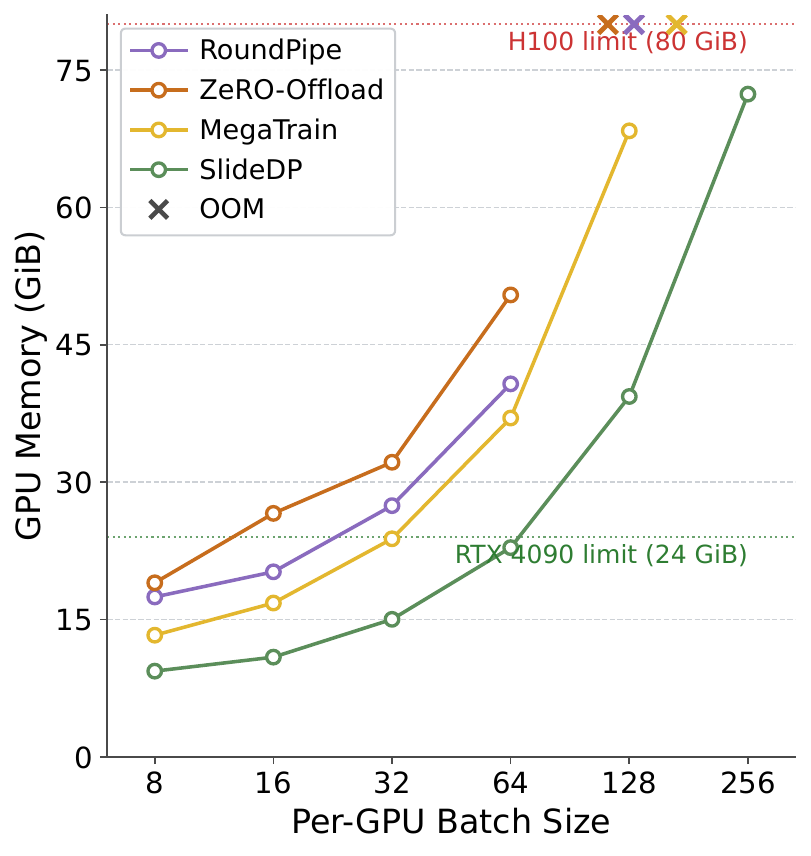}
        \caption{GPU memory vs. batch size for Qwen3-14B.}
        \label{fig:gpu-mem-batch-qwen3-14b}
    \end{minipage}
\end{figure*}

\fi

\section{Evaluation}
\label{sec:evaluation}

We first evaluate throughput and capacity, then isolate runtime mechanisms,
examine shared-host scaling, and assess policy gains and selection cost.
MoE workloads and loss comparisons assess model coverage and correctness.

\subsection{Experimental Setup}
\label{sec:eval-setup}

\textbf{Platforms.}
We evaluate \sys{} on three multi-GPU platforms. The PCIe-only
workstation has four RTX~4090 24GB GPUs, two Xeon Gold 6338N CPUs,
and 1~TiB DDR4 memory. The NVLink-bridged server has eight A800 80GB
PCIe GPUs, two Xeon Platinum 8358P CPUs, and 2~TiB DDR4 memory.
The NVSwitch server has eight H100 80GB SXM GPUs, two Xeon Platinum
8468 CPUs, and 2~TiB DDR5 memory; we use up to four H100s.
These platforms cover the three interconnect topologies in
Figure~\ref{fig:intro}(a--c).
Implementations use PyTorch Distributed, NCCL,
Transformers~\cite{wolf-etal-2020-transformers},
FlashAttention-2~\cite{dao2023flashattention2}, and compatible fused kernels
from Liger Kernel~\cite{hsu2024ligerkernelefficienttriton} and
Transformer Engine~\cite{TransformerEngine2024}.

\textbf{Workloads and baselines.}
We fine-tune all parameters with mixed precision~\cite{micikevicius2017mixed},
using BF16 computation and FP32 Adam states. Dense workloads span
Qwen3-1.7B through 32B~\cite{yang2025qwen3technicalreport} and
Qwen2.5-72B~\cite{qwen2025qwen25technicalreport}; the host-memory sweep
also includes Mistral-Small-24B~\cite{mistralai2025small24b}.
Sequence length defaults to 1024; batches are per GPU unless marked
global. Long-sequence runs use full-length inputs, all attention
positions enabled, and a position limit of 262,144.
Baselines are SlideFormer~\cite{yang2026slideformer},
ZeRO-Offload~\cite{ren2021zerooffload},
MegaTrain~\cite{yuan2026megatrain}, and
RoundPipe~\cite{luo2026roundpipe}.
SlideFormer's public multi-GPU implementation uses replicated parameter
delivery, CPU gradient reduction, and full-layer checkpointing with
activation offload; single-GPU points use its public single-GPU path.
ZeRO-Offload uses DeepSpeed ZeRO-3 with CPU parameter and optimizer
offload; RoundPipe uses synchronous mode.
Baselines use compatible fused kernels where available.
H100 comparisons also include GPU-resident
FSDP2~\cite{zhao2023pytorchfsdp,pytorch2026fsdp2docs} and Megatron-LM
with four-way tensor parallelism (TP4)~\cite{shoeybi2019megatron}.
Fixed-workload comparisons match global batch; TP4 realizes it through
resident microbatch and accumulation settings and plots global batch
divided by four. Model-size sweeps use each system's largest evaluated
feasible batch, annotated above each bar.

\textbf{Measurement protocol.}
Batch-scaling runs of \sys{} and \sys{}-fixed normally use 15 iterations, discarding the first eight as warmup. Each such point reports the arithmetic mean of
two runs. Throughput is measured after policy selection, whose overhead
is reported separately in Section~\ref{sec:eval-policy}.
Workload-specific measurement windows, repetition counts, and launch
configurations accompany the plotting data.

\textbf{Variants and metrics.}
In \sys{}-fixed, the runtime uses full-layer checkpointing,
activation offload, and fixed chunk sizes. \sys{} uses a selected policy
in the paired comparisons.
Throughput is global tokens per step divided by steady-state step time.
Model FLOPS equal this rate times the training FLOPs per token, computed
from actual projection dimensions with two operations per
multiply--accumulate. The count includes forward, backward, and
sequence-dependent attention, but excludes checkpoint replay,
optimizer work, and transfers.
GPU memory is CUDA peak reserved memory unless specified.
Host memory is proportional set size (PSS). For \sys{} and SlideFormer,
we sum sampled per-rank peaks; PSS apportions shared mappings across ranks.

\subsection{End-to-End Throughput}
\label{sec:eval-throughput}

\textbf{Batch-size scaling.}
We compare systems at matched batches
(Figures~\ref{fig:batch-4090-qwen3-8b} and~\ref{fig:batch-h100-qwen3-14b}).
On four RTX~4090 GPUs, \sys{} achieves
geometric-mean speedups of 1.83$\times$ over ZeRO-Offload
and 1.48$\times$ over MegaTrain across batches 4--32.
At batch 64, it reaches 9.81K tokens/s, while these baselines and
RoundPipe exhaust GPU memory.
On four H100s, the corresponding gains are 1.46$\times$ over ZeRO-Offload
(batches 16--64) and 1.59$\times$ over MegaTrain (batches 16--128).
At batch 256, \sys{} reaches 23.2K tokens/s and 1,048,576 tokens per step
without gradient accumulation. Its throughput is 1.41$\times$ and
1.47$\times$ the baselines' respective peaks in this sweep.

\begin{figure*}[t]
    \centering
    \begin{minipage}[t]{0.33\textwidth}
        \vspace{0pt}
        \centering
        \includegraphics[width=\linewidth]{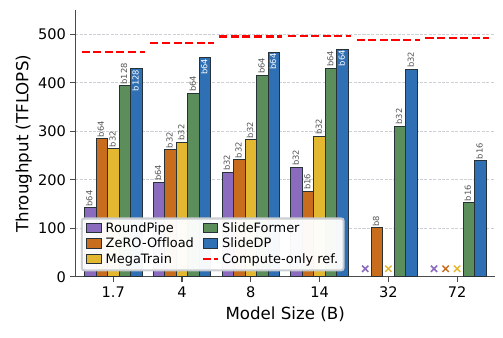}
        \caption{Model-size scaling throughput for Qwen models on 4$\times$RTX~4090.}
        \label{fig:model-scaling-4090}
    \end{minipage}
    \hfill
    \begin{minipage}[t]{0.33\textwidth}
        \vspace{0pt}
        \centering
        \includegraphics[width=\linewidth]{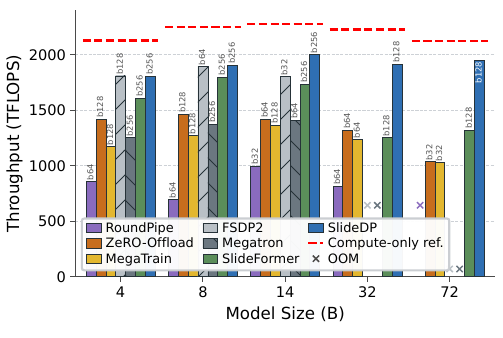}
        \caption{Model-size scaling throughput for Qwen models on 4$\times$H100.}
        \label{fig:model-scaling-h100}
    \end{minipage}
    \hfill
    \begin{minipage}[t]{0.33\textwidth}
        \vspace{0pt}
        \centering
        \includegraphics[width=\linewidth]{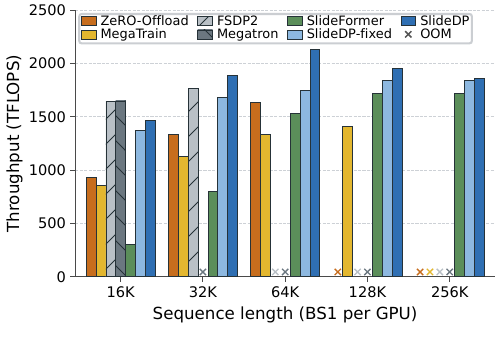}
        \caption{Sequence-length scaling for Qwen3-14B on 4$\times$H100.}
        \label{fig:seqlen-scaling-h100}
    \end{minipage}
\end{figure*}

\textbf{Model-size scaling.}
Figures~\ref{fig:model-scaling-4090} and~\ref{fig:model-scaling-h100}
compare model throughput at each system's largest evaluated feasible batch.
On RTX~4090, \sys{} sustains 429--470 TFLOPS across 1.7B--32B models;
Qwen3-32B achieves 4.24$\times$ ZeRO-Offload's throughput at batches 32
and 8, respectively. Qwen2.5-72B reaches 240 TFLOPS, 1.56$\times$
SlideFormer's throughput; the other three offloading baselines are
marked OOM.
On H100, \sys{} sustains 1.8--2.0 PFLOPS across 4B--72B models,
reaching 1.88$\times$ ZeRO-Offload's throughput at 72B with a fourfold
larger batch. At the plotted batches, it matches FSDP2 within 1\%
for 4B and 8B and exceeds it by 11.2\% for 14B (batches 256 and 32,
respectively).
The dashed H100 reference projects single-GPU layer/stage compute
measurements to four GPUs; FSDP2 and TP4 are measured end-to-end
GPU-resident comparisons.
These results show that layer streaming expands the feasible workload
range while maintaining competitive throughput against GPU-resident systems.

\textbf{Long sequences.}
Long sequences increase activation pressure and reduce available memory
headroom, making memory placement and pipeline coordination more
challenging.
Figure~\ref{fig:seqlen-scaling-h100} varies full attention length for
Qwen3-14B at one sequence per GPU. At 64K, \sys{} reaches 2.13 PFLOPS
versus 1.64 for ZeRO-Offload and 1.34 for MegaTrain (1.30$\times$ and
1.60$\times$). At 128K, it reaches 1.95 PFLOPS versus MegaTrain's
1.41 PFLOPS; it also completes 256K at 1.86 PFLOPS.
FSDP2 and TP4 are marked infeasible at 64K. Crosses denote measured
OOMs or longer sequences beyond a measured same-configuration OOM
boundary.
Together, these sweeps demonstrate matched-workload throughput gains
and a broader feasible workload range.

\subsection{Memory Efficiency and Capacity}
\label{sec:eval-memory}

With full-layer checkpointing and activation offload, \sys{} reduces
Qwen3-14B peak reserved memory by 51--59\% versus ZeRO-Offload and
29--42\% versus MegaTrain at common feasible batches
(Figure~\ref{fig:gpu-mem-batch-qwen3-14b}).
Batch 256 uses 72.4~GiB and is four times ZeRO-Offload's largest
feasible batch and twice MegaTrain's in this sweep.
The bounded window frees HBM for larger workloads and for retaining
activations to reduce transfers and replay (Section~\ref{sec:eval-policy}).

Figures~\ref{fig:batch-4090-qwen3-8b} and~\ref{fig:batch-h100-qwen3-14b}
report host PSS of 106--109~GiB for \sys{}, 183~GiB for ZeRO-Offload,
and 211--212~GiB for MegaTrain on RTX~4090 with Qwen3-8B at batches
4--32. On H100 with Qwen3-14B at batches 16--64, \sys{} uses about
202~GiB versus ZeRO-Offload's 342~GiB.
Shared master weights and optimizer states bound persistent host
storage, while GPU-retained activations avoid additional host copies.

\begin{figure}[t]
  \centering
  \includegraphics[width=\columnwidth]{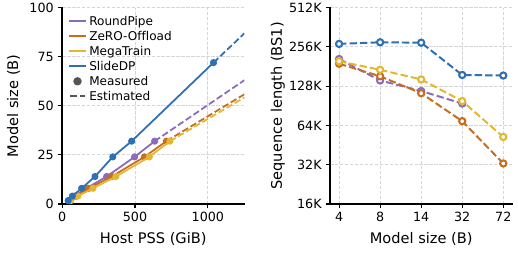}
  \caption{Host memory and estimated sequence capacity.}
  \label{fig:max-model-cpu}
\end{figure}

Figure~\ref{fig:max-model-cpu}(a) fits measured BS1 host footprints
through 32B: slopes are 14.2~GiB per billion parameters for \sys{}
and 19.8--22.9~GiB for other systems. Solid lines connect measurements;
dashed lines extrapolate these slopes. The lower growth rate supports
larger models within a fixed host-memory budget.
Panel (b) estimates sequence capacity from measured working sets at
one sequence per GPU, with 78~GiB per GPU and 1984~GiB host budgets.
It scales activation storage with token count and accounts for
RoundPipe's pipeline microbatches. MegaTrain uses sampled GPU memory
peaks; other systems use CUDA reserved peaks.
Projected 14B capacities are 276K tokens for \sys{}, 113K for
ZeRO-Offload, and 144K for MegaTrain;
Figure~\ref{fig:seqlen-scaling-h100} reports \sys{}'s successful 256K training run.

\begin{table*}[t]
\centering
\scriptsize
\caption{
AutoPolicy ablation across platforms and workloads. We report the change
from \sys{}-fixed to \sys{}. F-r1 and F-r0 denote full-layer
checkpointing with and without activation offload. S-r0 and SAC-r0 denote segment-level and
selective activation checkpointing without offload; N denotes no
checkpointing. Unlisted layers use F-r1.
}
\label{tab:autopolicy-ablation}
\setlength{\tabcolsep}{3.2pt}
\begin{tabular}{llcccccc}
\toprule
Platform / workload & Batch &
Base layout &
AutoPolicy layout &
Chunk \;(MiB) &
Throughput \;(Token/s) &
GPU mem. \;(GiB)&
CPU mem. \;(GiB)\\
\midrule
4$\times$RTX~4090, Qwen3-8B
& 16
& 36 F-r1
& 36 F-r0
& 64/64 $\rightarrow$ 8/64
& 6345 $\rightarrow$ 7198 \;(+13.4\%)
& 6.0 $\rightarrow$ 10.2
& 124.0 $\rightarrow$ 106.0 \;(-14.5\%) \\

4$\times$RTX~4090, Qwen3-8B
& 32
& 36 F-r1
& 30 F-r0 + 3 SAC-r0
& 64/64 $\rightarrow$ 8/16
& 9161 $\rightarrow$ 9415 \;(+2.8\%)
& 8.8 $\rightarrow$ 22.1
& 142.2 $\rightarrow$ 109.2 \;(-23.2\%) \\

4$\times$RTX~4090, Qwen3-8B
& 64
& 36 F-r1
& 31 F-r1 + 5 SAC-r0
& 64/64 $\rightarrow$ 8/16
& 9684 $\rightarrow$ 9740 \;(+0.6\%)
& 14.5 $\rightarrow$ 22.0
& 178.0 $\rightarrow$ 168.3 \;(-5.5\%) \\

\midrule
4$\times$H100, Qwen3-14B
& 32
& 40 F-r1
& 31 F-r0 + 9 N
& 64/64 $\rightarrow$ 128/128
& 17370 $\rightarrow$ 19530 \;(+12.4\%)
& 15.0 $\rightarrow$ 76.6
& 261.9 $\rightarrow$ 200.6 \;(-23.4\%) \\

4$\times$H100, Qwen3-14B
& 64
& 40 F-r1
& 37 F-r0 + 3 SAC-r0
& 64/64 $\rightarrow$ 32/128
& 21658 $\rightarrow$ 22635 \;(+4.5\%)
& 22.8 $\rightarrow$ 76.4
& 341.6 $\rightarrow$ 200.6 \;(-41.3\%) \\

4$\times$H100, Qwen3-14B
& 128
& 40 F-r1
& 15 F-r1 + 25 F-r0
& 64/64 $\rightarrow$ 8/128
& 22033 $\rightarrow$ 22657 \;(+2.8\%)
& 39.4 $\rightarrow$ 75.9
& 501.6 $\rightarrow$ 313.1 \;(-37.6\%) \\

\midrule
4$\times$A800, Qwen3-4B
& 64
& 36 F-r1
& 10 F-r0 + 10 N
& 64/64 $\rightarrow$ 32/128
& 26805 $\rightarrow$ 28964 \;(+8.1\%)
& 11.1 $\rightarrow$ 77.3
& 131.7 $\rightarrow$ 94.9 \;(-28.0\%) \\

4$\times$A800, Qwen3-8B
& 64
& 36 F-r1
& 29 F-r1 + 7 N
& 64/64 $\rightarrow$ 128/64
& 15737 $\rightarrow$ 16281 \;(+3.5\%)
& 14.9 $\rightarrow$ 69.0
& 179.6 $\rightarrow$ 167.3 \;(-6.8\%) \\

4$\times$A800, Qwen3-32B
& 64
& 64 F-r1
& 13 F-r0 + 3 N
& 64/64 $\rightarrow$ 32/128
& 3848 $\rightarrow$ 3923 \;(+1.9\%)
& 26.5 $\rightarrow$ 71.8
& 638.7 $\rightarrow$ 585.2 \;(-8.4\%) \\

\bottomrule
\end{tabular}%
\end{table*}

\subsection{Shared-Host Runtime Mechanisms}
\label{sec:eval-runtime}
\label{sec:eval-breakdown}

\textbf{Reducing shared-host demand.}
The fixed-policy comparisons in Figures~\ref{fig:batch-4090-qwen3-8b}
and~\ref{fig:batch-h100-qwen3-14b} evaluate the shared-host datapath.
On RTX~4090, both paths use replicated parameter delivery and full-layer
checkpointing with activation offload. \sys{} aggregates gradients on
GPU before their return, replacing per-rank D2H transfers and CPU
aggregation with one aggregated gradient per layer. On H100, the fixed
runtime also uses sharded delivery over NVSwitch. These changes reduce
shared-host service demand while retaining the same activation policy.
At batches 4--16 on RTX~4090, \sys{}-fixed achieves 2.32--2.54$\times$
SlideFormer's throughput; at batches 16--32 on H100, it achieves
4.03--4.18$\times$. Larger batches provide more GPU computation to hide
host work, narrowing the gap as described by the exposure model
(Section~\ref{sec:step-time-model}).

\begin{figure}[h]
  \centering
  \includegraphics[width=\columnwidth]{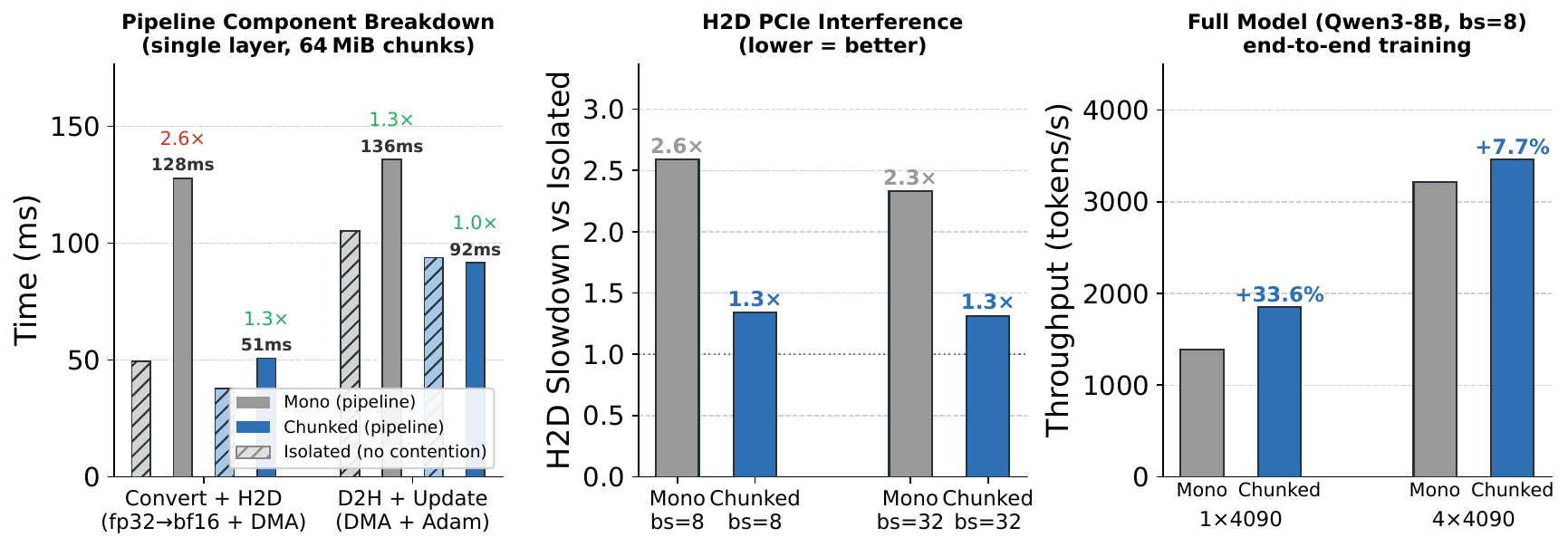}
    \caption{Chunked overlap: layer microbenchmarks and Qwen3-8B training on one and four RTX~4090 GPUs.}
    \label{fig:pipeline-breakdown}
\end{figure}

\textbf{Chunked pipeline overlap.}
After reducing shared-host demand, \sys{} uses chunking to overlap the
remaining transfers and updates (Section~\ref{sec:pipeline}).
Figure~\ref{fig:pipeline-breakdown} evaluates this overlap within a layer.
In the single-layer experiment with 64~MiB
chunks, conversion plus H2D takes 50.8~ms under pipeline contention,
down from 127.8~ms with whole-layer transfers. D2H plus CPU update
falls from 135.9 to 91.7~ms. Across the two computation windows in
the middle panel, chunking reduces H2D slowdown relative to isolated
execution from 2.3--2.6$\times$ to approximately 1.3$\times$.
Shorter service paths advance parameter and update readiness, reducing
exposed waits in the layer pipeline (Section~\ref{sec:step-time-model}).

The right panel evaluates Qwen3-8B at batch eight per GPU, using full-length
inputs on four GPUs and padded variable-length inputs on one.
Chunking increases throughput from 3,218 to 3,465 tokens/s on four
RTX~4090s (+7.7\%) and from 1,386 to 1,852 tokens/s on one (+33.6\%),
averaging two runs per configuration.
Each pair keeps layer-wise prefetching and CPU allocation fixed.
The four-GPU gain demonstrates that chunk-level overlap improves
end-to-end execution with multiple ranks sharing host resources.

\subsection{Understanding Shared-Host Scaling}
\label{sec:eval-scaling}
\label{sec:eval-topology}

The computation-window effect also appears in the sequence sweep
(Figure~\ref{fig:seqlen-scaling-h100}): longer sequences narrow the gap
between \sys{}-fixed and SlideFormer. We next examine the model in
Section~\ref{sec:step-time-model} by varying GPU count and delivery routes.

\begin{figure}[h]
  \centering
  \includegraphics[width=\columnwidth]{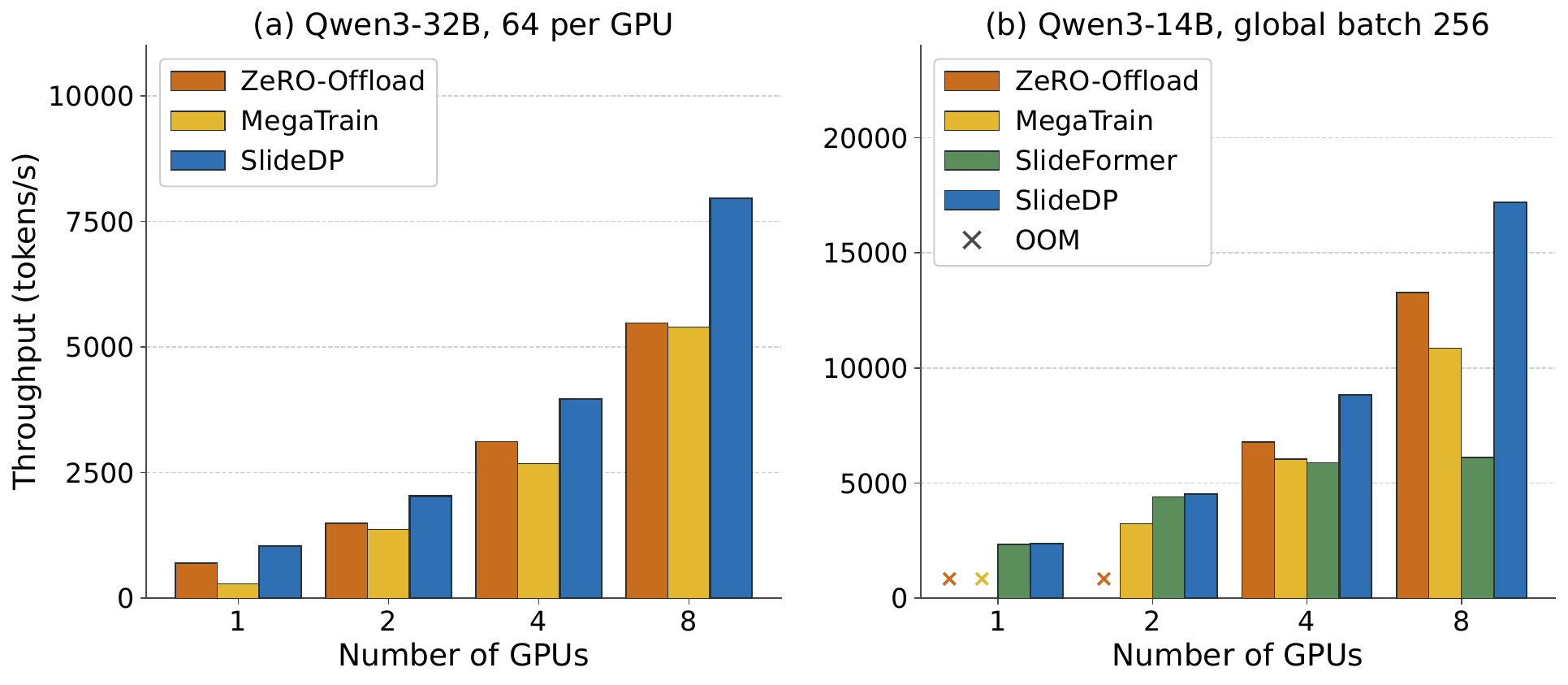}
    \caption{A800 scaling: (a) Qwen3-32B, 64 sequences per GPU; (b) Qwen3-14B, global batch 256.}
    \label{fig:gpu-scaling-a800}
\end{figure}

\textbf{GPU count and computation windows.}
Figure~\ref{fig:gpu-scaling-a800} separates weak and strong scaling.
For Qwen3-32B at 64 sequences per GPU, \sys{} reaches 7,963 tokens/s
on eight A800s, versus 5,478 for ZeRO-Offload and 5,396 for MegaTrain.
Parallel efficiency of 95--98\% across two, four, and eight GPUs
indicates near-linear weak scaling in the GPU-bound regime.
For Qwen3-14B at global batch 256, \sys{}-fixed achieves 1.94$\times$
scaling from four to eight GPUs (8.84K to 17.18K tokens/s), versus
1.04$\times$ for SlideFormer.

Figure~\ref{fig:intro}(e) reports excess step time over the matched
compute reference as an aggregate indicator of exposed execution cost.
At eight A800s and global batch 256, the excess is 0.90~s for \sys{},
28.57~s for SlideFormer, and 9.77~s for MegaTrain.
Low exposed cost lets the fixed runtime sustain scaling as each rank's
computation window shrinks.
At fixed local batch, \sys{} keeps excess time at roughly 1--2~s across
one to eight GPUs (Figure~\ref{fig:intro}(d)).

\begin{figure}[h]
  \centering
  \includegraphics[width=\columnwidth]{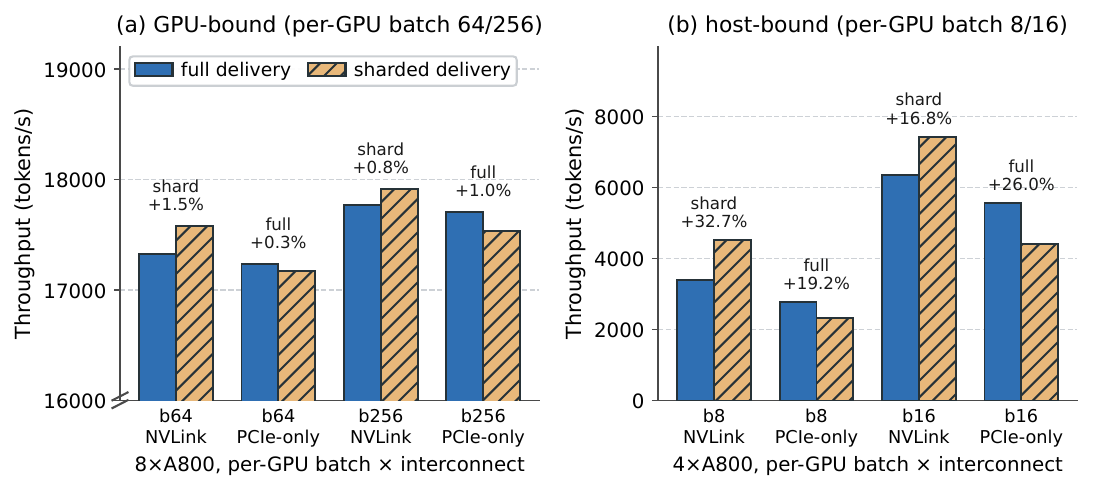}
    \caption{Qwen3-14B route sensitivity on A800 with and without NVLink bridges.}
    \label{fig:topology-sensitivity}
\end{figure}

\textbf{Topology and communication routes.}
Figure~\ref{fig:topology-sensitivity} compares replicated and sharded
parameter delivery for Qwen3-14B on A800.
In host-bound four-GPU workloads, sharded delivery improves throughput
by 16.8--32.7\% with NVLink bridges; after their removal, replicated
delivery improves it by 19.2--26.0\%.
Sharding trades repeated H2D for GPU-side reconstruction, whereas
replication avoids reconstruction on a slower GPU path.
In GPU-bound eight-GPU workloads, all route/topology combinations
remain within 2.5\% at each batch.
Route choice therefore depends jointly on topology and the computation
available to hide communication.

\subsection{Measurement-Guided Policy Selection}
\label{sec:eval-policy}

We compare \sys{} with \sys{}-fixed under the same shared-host runtime
to measure the combined benefit of selecting routes, chunk sizes, and
activation layouts (Section~\ref{sec:autopolicy}).
Routes and chunks are probed first; shortlisted activation candidates are
then calibrated in a reduced pipeline before budgeted allocation,
validation, and bounded refinement.
Table~\ref{tab:autopolicy-ablation} compares selected chunks and layouts
with matched fixed controls in a separate measurement series from the
batch-scaling curves.

\textbf{Selected layouts and gains.}
In Figure~\ref{fig:batch-h100-qwen3-14b}, policy selection raises
Qwen3-14B throughput on H100 at batch 16 from 10,231 to 12,832 tokens/s
(+25.4\%). At 64K, policy selection gains 21.7\% over \sys{}-fixed
(Figure~\ref{fig:seqlen-scaling-h100}).

Table~\ref{tab:autopolicy-ablation} illustrates how the selected layouts
use HBM to reduce different costs. On RTX~4090 at batch 16, retaining
all 36 layer inputs removes activation offload/reload traffic; together
with 8/64~MiB H2D/D2H chunks, the configuration gains 13.4\% throughput
and reduces host PSS by 14.5\%.
On H100 at batch 32, combining 31 F-r0 layers with nine no-checkpoint
layers eliminates replay in the latter. This configuration uses 76.6~GiB
versus 15.0~GiB for the fixed policy and gains 12.4\% throughput,
turning the bounded window's memory headroom into faster execution.

Larger batches use mixed layouts as activation demand grows. On H100
at batch 128, 25 F-r0 layers retain their inputs while 15 F-r1 layers
offload them; the configuration gains 2.8\% throughput and reduces host
PSS by 37.6\%. On RTX~4090 at batch 64, five SAC-r0 layers retain
selected intermediates alongside 31 F-r1 layers, yielding a 0.6\% gain
and 5.5\% less host PSS.
On A800 at batch 64, selected configurations gain
1.9--8.1\% throughput with 6.8--28.0\% less host PSS under fixed
sharded delivery and GPU reduction.

\textbf{Activation placement.}
Each checkpoint family combines with an offload ratio
(Section~\ref{sec:elastic-checkpointing}).
The selected layouts offload compact full-layer inputs while retaining
larger intermediate sets from finer-grained policies on GPU.
Retaining these intermediates reduces replay without adding activation
traffic to the PCIe paths used by parameters and gradients.

\textbf{Selection overhead and amortization.}
On four H100s with Qwen3-14B and 32 sequences per GPU, a separate
selection run reuses a cached single-layer activation profile and takes
1218~s (20.3~min) for route/chunk probes, pipeline calibration, and final
validation. It selects 40 F-r0 layers and 32/128~MiB H2D/D2H chunks.
With routes and chunks fixed, its final validation reduces step time
from 7.797~s with F-r1 to 6.828~s with the selected layout.
The measured 0.970~s saving per step yields an estimated break-even
of 1,257 steps (2.72~hours of baseline training). This one-time cost
would account for 1.4\% of a 24-hour training budget.

\subsection{MoE Training and Correctness}
\label{sec:eval-moe}

\begin{table}[t]
  \centering
  \caption{MoE throughput on four H100 GPUs (tokens/s).}
  \label{tab:moe-throughput}
  \begingroup
  \footnotesize
  \setlength{\tabcolsep}{3pt}
  \renewcommand{\arraystretch}{1.1}
  \begin{tabular*}{\columnwidth}{@{\extracolsep{\fill}}lrrr@{}}
    \toprule
    Model / per-GPU $B\times S$
      & \textcolor[HTML]{2F6FB3}{\rule{4pt}{4pt}}\,\sys{}
      & \textcolor[HTML]{C76D1D}{\rule{4pt}{4pt}}\,ZeRO-Offload
      & \textcolor[HTML]{E3B72F}{\rule{4pt}{4pt}}\,MegaTrain \\
    \midrule
    \shortstack[l]{Qwen3.6-35B-A3B\\$32\times4096$} & 44,715 & 13,931 & 26,102 \\
\addlinespace[2pt]
\shortstack[l]{Gemma4-26B-A4B\\$64\times1024$} & 27,347 & 13,822 & 18,993 \\
\bottomrule
  \end{tabular*}%
  \endgroup
\end{table}

The shared-host DP runtime also supports MoE layers.
Table~\ref{tab:moe-throughput} compares
Qwen3.6-35B-A3B~\cite{qwen2026qwen36model} and
Gemma4-26B-A4B~\cite{google2026gemma4model} on four H100 GPUs at matched
batch sizes and sequence lengths across systems. \sys{} reaches 44.7K
and 27.3K tokens/s, respectively. It delivers $1.98$--$3.21\times$
ZeRO-Offload's and $1.44$--$1.71\times$ MegaTrain's throughput.

\textbf{Correctness.}
We assess dense-model correctness by comparing \sys{} with
GPU-resident FSDP2 over 200 training steps of Qwen3-14B on four H100s,
using a per-GPU batch size of 16 and sequence length 1024.
We match the initial weights, C4 samples~\cite{raffel2020t5} and their
order, and optimizer settings.
All ranks complete with finite loss; the maximum absolute differences
in global-batch mean loss are $5.2\times10^{-4}$ for F-r1 and
$6.3\times10^{-4}$ for the selected layout; the loss curves followed consistent trajectories.
The close agreement with FSDP2 in BF16 training loss supports the
correctness of \sys{}'s synchronous execution
(Section~\ref{sec:pipeline}).

\section{Conclusion}
\label{sec:conclusion}

This paper presented \sys{}, a shared-host runtime for full-parameter LLM fine-tuning with synchronous data parallelism. Shared state and cross-rank pipelining reduce redundant host work and overlap the remaining transfers and updates with GPU computation. The analytical model explains when reducing shared-resource demand shortens a step; measurements guide communication and activation policies for each workload and topology. The broader design lesson is to manage communication and GPU memory according to their effects on exposed pipeline work. This coordination translates host-memory capacity into larger feasible training workloads while sustaining throughput across commodity workstations and multi-GPU servers.

\FloatBarrier
\bibliographystyle{ACM-Reference-Format}
\bibliography{sample-base}
\end{document}